\documentclass[useAMS,usenatbib]{mnras}

\pdfoutput=0

\usepackage{amssymb}
\usepackage{amsmath}
\usepackage{times}
\usepackage{graphicx}
\usepackage{wasysym}

\def\aap{A\&A}
\def\apj{ApJ}

\def\apjl{ApJ}
\def\pasa{PASA}
\def\mnras{MNRAS}
\def\araa{ARA\&A}
\def\aj{AJ}

\def\apjs{ApJS}
\def\pasp{PASP}
\def\rmxaa{Rmxaa}

\newcommand{\mincir}{\raise
-2.truept\hbox{\rlap{\hbox{$\sim$}}\raise5.truept 
\hbox{$<$}\ }}
\newcommand{\magcir}{\raise
-2.truept\hbox{\rlap{\hbox{$\sim$}}\raise5.truept
\hbox{$>$}\ }}
\newcommand{\minmag}{\raise-2.truept\hbox{\rlap{\hbox{$<$}}\raise
6.truept\hbox
{$>$}\ }}

\newcommand{\be}{\begin{equation}}
\newcommand{\ee}{\end{equation}}
\newcommand{\ba}{\begin{eqnarray}}
\newcommand{\ea}{\end{eqnarray}}
\newcommand{\brr}{\begin{array}}
 
\newcommand{\err}{\end{array}}
\newcommand{\bc}{\begin{center}}
\newcommand{\ec}{\end{center}}

\DeclareMathAlphabet{\mathsc}{OT1}{cmr}{m}{sc}
\def\testbx{bx}%
\DeclareRobustCommand{\ion}[2]{%
\relax\ifmmode
\ifx\testbx\f@series
{\mathbf{#1\,\mathsc{#2}}}\else
{\mathrm{#1\,\mathsc{#2}}}\fi
\else\textup{#1\,{\mdseries\textsc{#2}}}%
\fi}

\title[Systematic uncertainties in the SFR-M*]{The high redshift SFR-M* relation is sensitive to the employed star formation rate and stellar mass indicators: Towards addressing the tension between observations and simulations.}

\author[A. Katsianis et al.]{A. Katsianis$^{1}$ $^{2}$ $^{3}$\thanks{E-mail:
    kataunichile@gmail.com, kata@sjtu.edu.cn}, V Gonzalez$^{4}$ $^{5}$, D. Barrientos$^{3}$, X. Yang $^{1}$ $^{2}$, C.D.P. Lagos$^{6}$ $^{7}$ $^{8}$, \newauthor J. Schaye$^{9}$, P. Camps$^{10}$, A. Tr\v{c}ka$^{10}$, M. Baes$^{10}$, M. Stalevski$^{11}$ $^{10}$, G.A. Blanc$^{12}$ $^{3}$ and T. Theuns$^{13}$
  \\ $^1$ Tsung-Dao Lee Institute, Shanghai Jiao Tong University, Shanghai 200240, China \\
    $^2$  Department of Astronomy, Shanghai Key Laboratory for Particle Physics and Cosmology, Shanghai Jiao Tong University, Shanghai 200240, China \\
  $^3$ Department of Astronomy, Universitad de Chile, Camino El Observatorio 1515, Las Condes, Santiago, Chile \\
  $^4$ Chinese Academy of Sciences South America Center for Astronomy, China-Chile Joint Center for Astronomy, Camino del Observatorio 1515, Las Condes, Chile \\
  $^5$ Centro de Astrofísica y Tecnologías Afines (CATA), Camino del Observatorio 1515, Las Condes, Santiago, Chile \\
  $^{6}$ International Centre for Radio Astronomy Research (ICRAR), M468, University of Western Australia, 35 Stirling Hwy, Crawley, WA 6009, Australia.\\
  $^{7}$ ARC Centre of Excellence for All Sky Astrophysics in 3 Dimensions (ASTRO 3D).\\
  $^{8}$ Cosmic Dawn Center (DAWN) \\
    $^{9}$ Leiden Observatory, Leiden University, PO Box 9513, NL-230 0 RA Leiden, The Netherlands \\
  $^{10}$ Sterrenkundig Observatorium, Universiteit Gent, Krijgslaan 281, B-9000 Gent, Belgium \\
  $^{11}$ Astronomical Observatory, Volgina 7, 11060 Belgrade, Serbia \\
  $^{12}$ Observatories of the Carnegie Institution for Science, 813 Santa Barbara St, Pasadena, CA, 91101, USA \\
  $^{13}$ Institute for Computational Cosmology, Department of Physics, University of Durham, South Road, Durham, DH1 3LE, UK} 
   
\begin{document}

\maketitle

\begin{abstract}

  There is a severe tension between the observed star formation rate (SFR) - stellar mass (${\rm M}_{\star}$) relations reported by different authors at $z = 1-4$. In addition, the observations have not been successfully reproduced by state-of-the-art cosmological simulations which tend to predict a factor of 2-4 smaller SFRs at a fixed ${\rm M}_{\star}$. We examine the evolution of the SFR$-{\rm M}_{\star}$ relation of $z = 1-4 $ galaxies using the SKIRT simulated spectral energy distributions of galaxies sampled from the EAGLE simulations. We derive SFRs and stellar masses by mimicking different observational techniques. We find that the tension between observed and simulated SFR$-{\rm M}_{\star}$ relations is largely alleviated if similar methods are used to infer the galaxy properties. We find that relations relying on infrared wavelengths (e.g. 24 ${\rm \mu m}$,  MIPS - 24, 70 and 160 ${\rm \mu m}$ or SPIRE - 250, 350, 500 ${\rm \mu m}$) have SFRs that exceed the intrinsic relation by 0.5 dex. Relations that rely on the spectral energy distribution fitting technique underpredict the SFRs at a fixed stellar mass by -0.5 dex at $z \sim 4$ but overpredict the measurements by 0.3 dex at $z \sim 1$. Relations relying on dust-corrected rest-frame UV luminosities, are flatter since they overpredict/underpredict SFRs for low/high star forming objects and yield deviations from the intrinsic relation from 0.10 dex to -0.13 dex at $z \sim 4$. We suggest that the severe tension between different observational studies can be broadly explained by the fact that different groups employ different techniques to infer their SFRs.
  
\end{abstract}


\section{Introduction}

Star formation rate (${\rm SFR}$) and stellar mass (${\rm M_{\star}}$) are two fundamental properties of galaxies, since each can provide a useful census for galaxy formation and evolution. The ${\rm SFR}$-${\rm M_{\star}}$ plane can be  loosely separated into three different Gaussian distributions \citep{Bisigello2018}, corresponding to 1) the quenched/passive galaxies, 2) the star forming galaxies, and 3) the starburst galaxies. A range of observational studies have exhibited the existence of a relation between star formation rate and stellar mass (${\rm M_{\star}}$) for $z \simeq 0-8$, especially for the star forming population \citep{Noeske2007,Elbaz2011,Whitaker2014,Tomczak2016,Popesso2019,Davies2019,Katsianis2019}, to the extent that such correlation has been labelled as the Main Sequence (MS)\footnote{ In order to select star forming galaxies and define the MS, different authors use different criteria (e.g. minimum threshold for of sSFR = SFR/${\rm M_{\star}}$, UVJ color-color selection, ridge line in the 3D surface defined by the SFR-mass-number density relation) which should ideally remove galaxies with low specific star formation rates from their ``parent'' samples. However, the thresholds differ significantly in value from one study to an other \citep{Renzini2015} making the comparison between the results of different authors challenging.}. Samples with no selection of star forming galaxies produce either flatter or ``bending'' SFR-$M_{\star}$ relations at low redshifts ($z < 1$) and higher masses \citep{Drory08,Bauer11,Bisigello2018} due to the presence of the quenched population, which contains galaxies with lower star formation rates at a fixed stellar mass.

In order to retrieve the intrinsic properties of galaxies and determine the ${\rm SFR}$-${\rm M_{\star}}$ relation, different observational studies rely on  different models and SFR/${\rm M_{\star}}$ diagnostics. Stellar masses are typically calculated via the Spectral Energy Distribution (SED) fitting technique \citep[e.g.][]{Kriek2009,Conroy2013,Boquien2019}, for which various assumptions are required (e.g. initial mass function, star formation history, dust attenuation model, metallicity fraction). Furthermore, different studies employ different calibrations/wavelengths in order to derive galaxy SFRs like IR$_{24 \mu m}$ luminosities \citep{Rodighiero10,Guo2013,Whitaker2014,Guo2015}, H$\alpha$ luminosities \citep{Sanchez2018,Cano2019}, the SED fitting technique \citep{Drory08,Kajisawa2010,Karim2011,Bauer11,deBarros,Kurczynski2017} or UV luminosities \citep{Salim2007,bouwens2012,Santini2017,Blanc2019}. A number of questions arise. The different diagnostics, assumptions and methodologies used by different observational studies produce results that are in agreement ? If not, is there a way to decipher the effect of the assumed methodology ?
  
In the last years an increasing number of authors have reported a discrepancy between the SFRs inferred by different methodologies \citep{Utomo2014,Fumagalli2014,Boquien2014,Hayward2014,Davies2016,Davies2017,Katsianis2017}. In addition, \citet{Katsianis2015} demonstrated that there is a severe tension of $\simeq 0.2-1$ dex between the  observed ${\rm SFR}$-${\rm M_{\star}}$ relations at $z \sim 1-4$ reported by different groups and suggested that the lack of consensus between different authors has its roots in the diversity of techniques used in the literature to estimate SFRs and also in sample selection effects. Furthermore, \citet{Davies2016} pointed out that different methods yield relations with inconsistent slopes and normalizations. In addition, \citet{Speagle14} and \citet{Renzini2015} suggested that the logarithmic slope $\alpha$ of the MS relation, which can be fitted by $ {\rm Log_{10}(SFR)} = {\rm \alpha Log_{10}M_{\star} + c} $, ranges from $\sim$ 0.4 up to $\sim$ 1.0 from study to study, while the normalization $ c $ differs from -8.30 up to -1.80 at redshift $z \sim 2.0$. Some authors find significant evolution for the slope ($\alpha (z)$ = $0.70-0.13 z$) at $z \sim 0-2.5$ \citep{Whitaker2012}, while others indicate no evolution \citep{Dunne2009,Karim2011}. The scatter of the relation also varies in the literature. Some authors report that $\sigma_{SFR}$ is constant with stellar mass and redshift \citep{Rodighiero10,Schreiber2015} while others suggest that the dispersion is mass/redshift dependent \citep{Guo2013,Katsianis2019}.

Cosmological hydrodynamic simulations from different collaborations such as EAGLE \citep{Schaye2015,Crain2015}, Illustris \citep{Vogelsberger2014}, IllustrisTNG \citep{Pillepich2018} and ANGUS \citep{TescariKaW2013,Katsianis2014}, have successefully replicated a range of observables and thus can provide information about the ${\rm SFR}$-${\rm M_{\star}}$ relation. However, the simulations have not been able to reproduce most of the observed ${\rm SFR}$-${\rm M_{\star}}$ relations reported in the literature. Indeed most groups report tension with observations, especially at $z \simeq 1-2$ \citep{Sparre2014,Furlong2014,Katsianis2015,Donnari2019}. The questions that arise are: Why cosmological hydrodynamic simulations have been unable to reproduce most of the observed  ${\rm SFR}$-${\rm M_{\star}}$ relations at high redshifts ? Can they provide insights on the tension between different observational studies ? 

Evaluating the determination of galaxy properties from different methodologies requires a galaxy sample with known intrinsic properties. Thus, a range of articles have examined separately the recovery of stellar masses \citep{Wuyts2009,Hayward2015,Torray2015,Camps2016,Price2017} and SFRs \citep{Kitzbichler2007,Maraston2010,Pforr2012} using mock/simulated galaxies. Hence, mock surveys \citep{Snyder2011,Camps2018,Liang2019}, which involve objects with known SFRs, stellar masses and fluxes at various key bands (e.g. GALEX-FUV, SDSS-u, 2MASS-Ks, WISE 3.4 $\mu$m or Spitzer 24 $\mu$m), are ideal to explore the effect of SFR and ${\rm M_{\star}}$ diagnostics on the inferkatsianis antoniosred ${\rm SFR}$-${\rm M_{\star}}$ relation.

In this paper we employ the mock SEDs described in \citet{Camps2018} and derive properties following observational methodologies used in the literature. We derive stellar masses through the SED fitting technique \citep{Kriek2009}. SFRs are calculated using the 24, 70 and 160 ${\rm \mu m}$ luminosities and their relation with the Total IR (TIR) luminosity \citep{DaleHelou2002,Wuyts2008}, fitting the SPIRE 250, 350 and 500 ${\rm \mu m}$ fluxes to the \citet{Dale2014} templates, dust-corrected UV luminosities via the IRX-$ \beta $ relation \citep{meurer1999} and the SED fitting technique. The analysis allows us to address the discrepancy between different observational methodologies to infer SFRs and stellar masses while it provides insights on the tension between cosmological hydrodynamic simulations and observational studies at high redshifts. In section \ref{thecode} we present a comparison between a range of observed relations and EAGLE simulations. In section \ref{thecode2} we briefly present the EAGLE+SKIRT data while in subsection \ref{thecode11} we describe the methodologies used to derive SFRs and stellar masses from the simulated galaxies. In section \ref{thecode3} we perform the comparison between observations and simulations. In section \ref{Disc} we draw our conclusions. In the appendix \ref{Appendix} we provide a comparison between the inferred and intrinsic star formation rates and stellar masses.

\begin{figure*}
\centering
\includegraphics[scale=0.37]{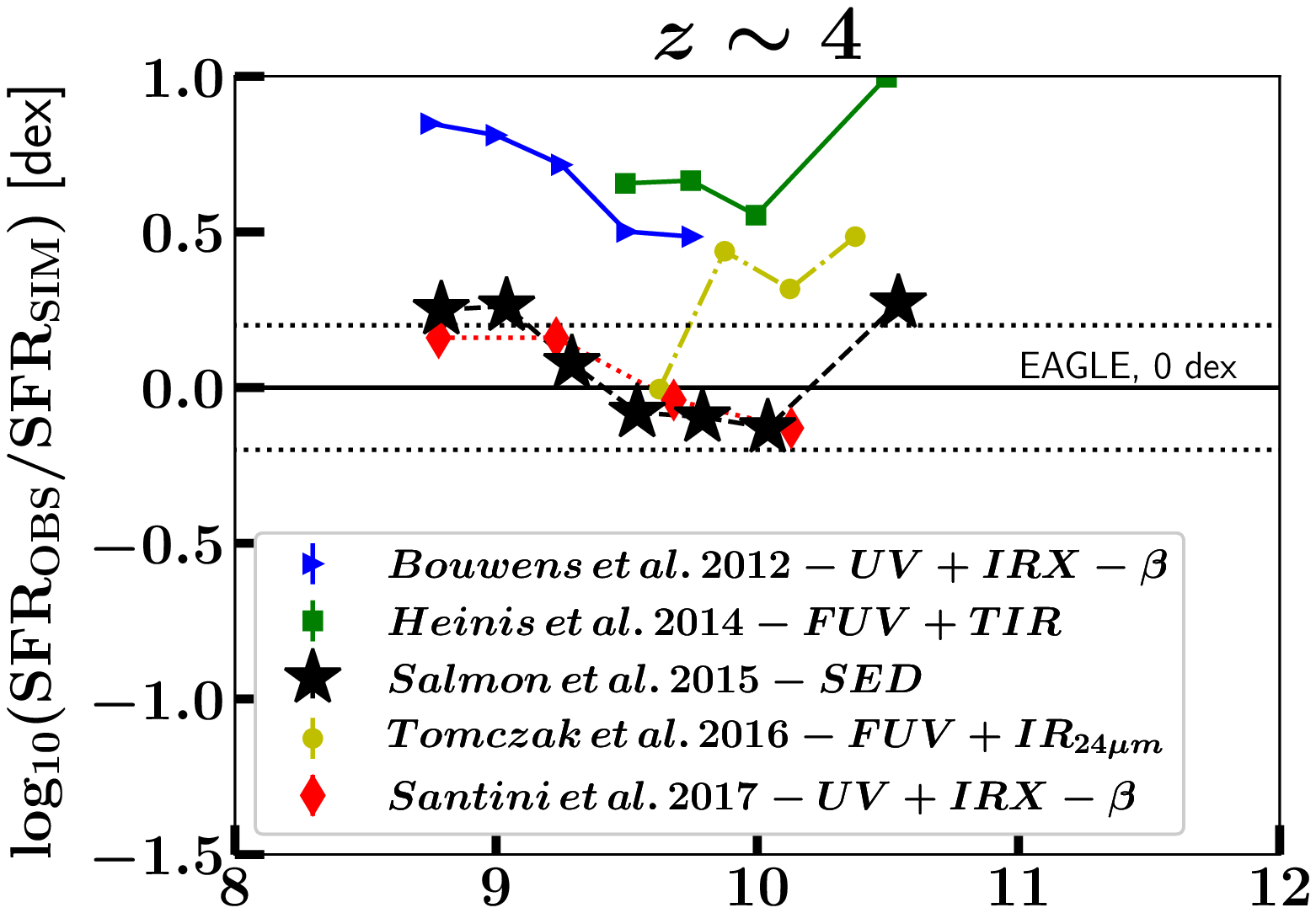} 
\includegraphics[scale=0.37]{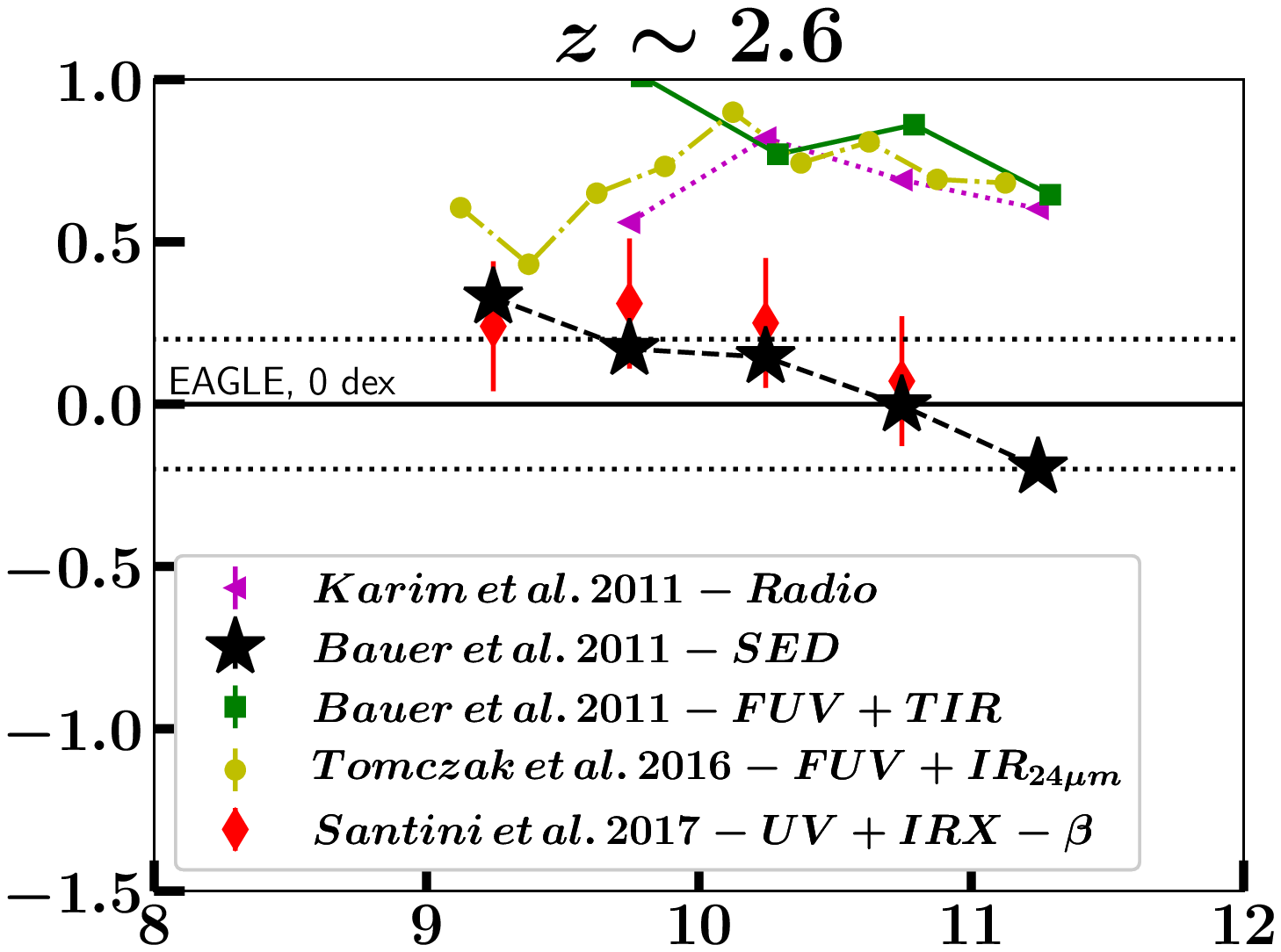} 
\includegraphics[scale=0.37]{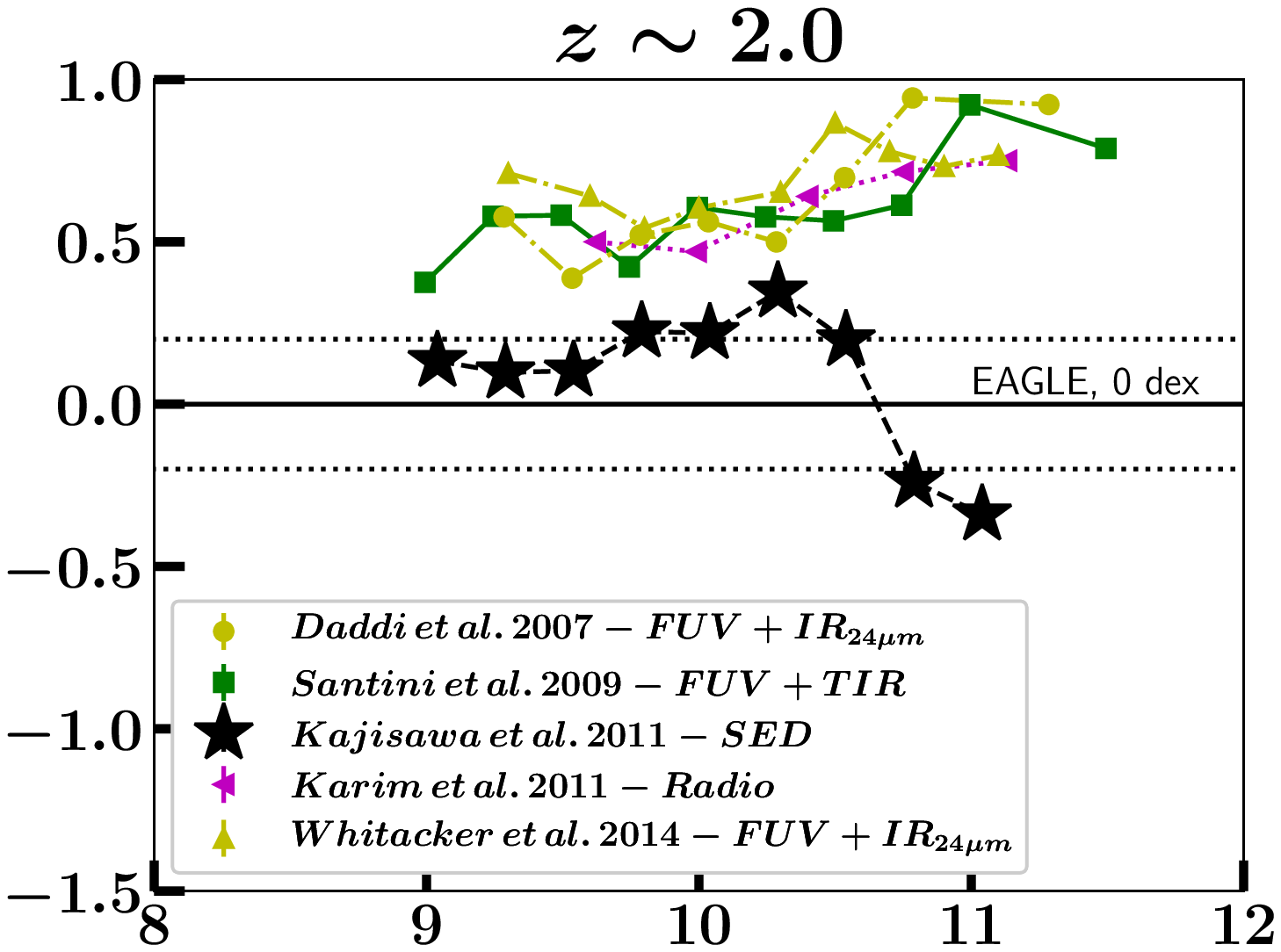}
\includegraphics[scale=0.37]{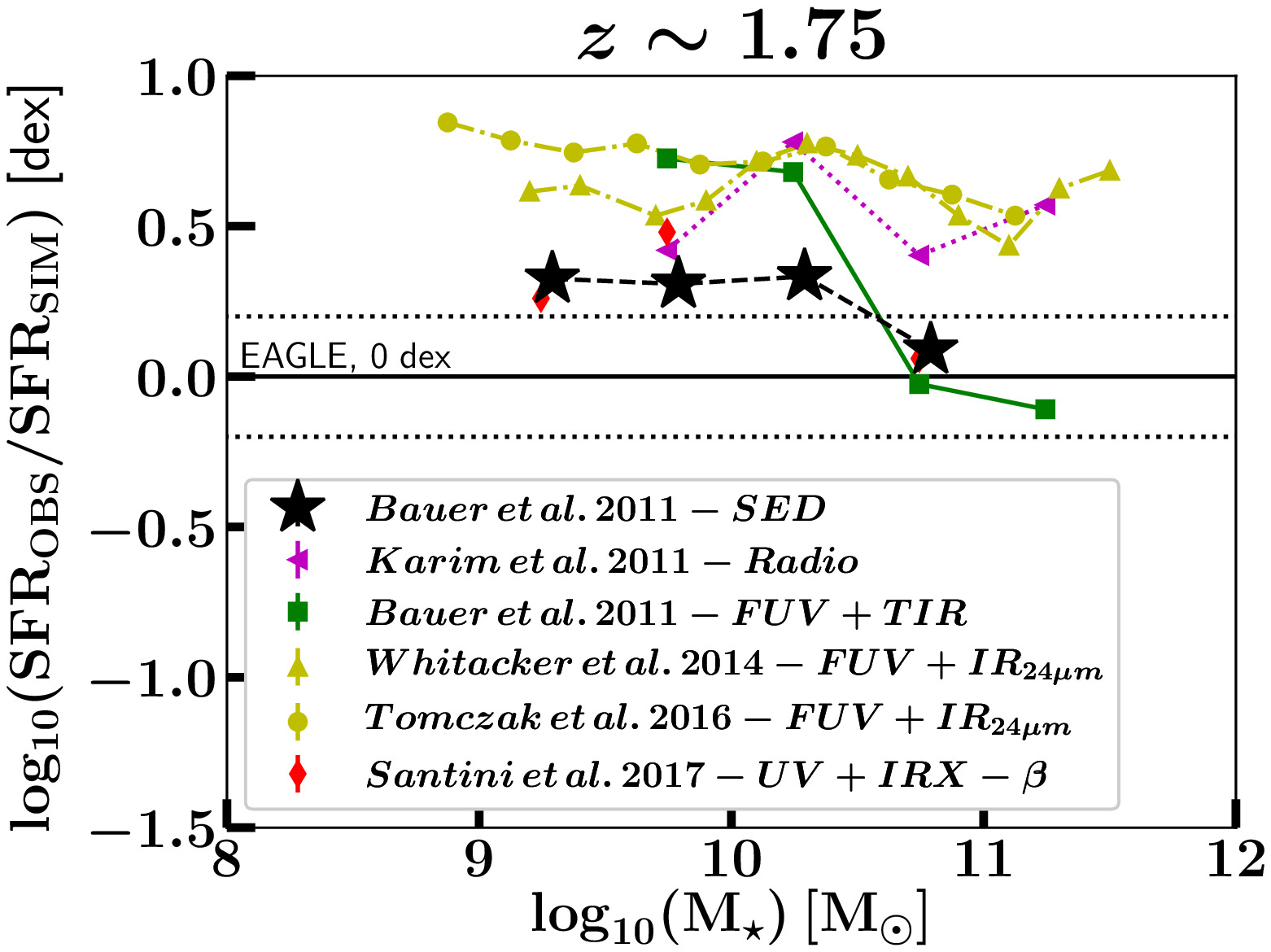} 
\includegraphics[scale=0.37]{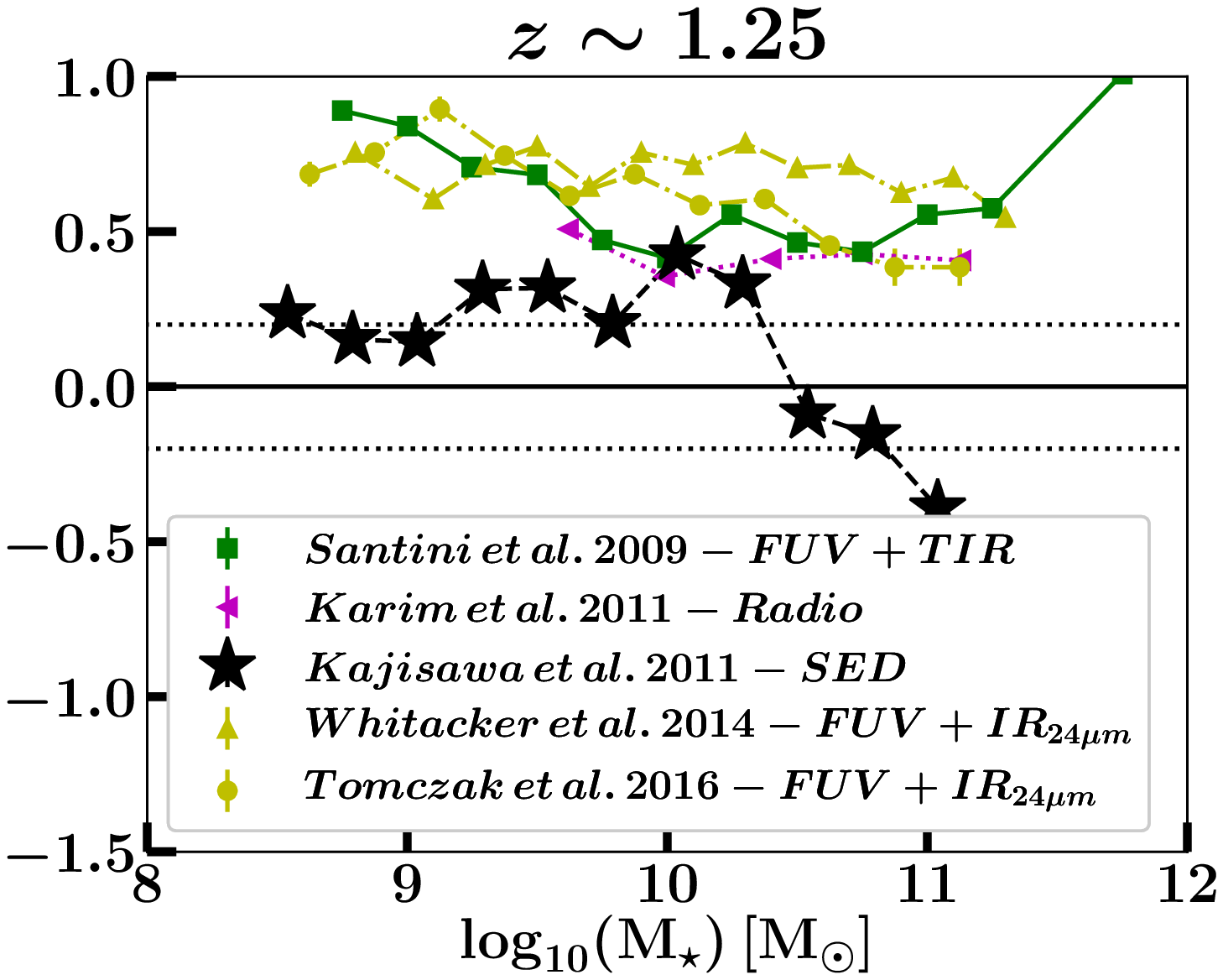} 
\includegraphics[scale=0.37]{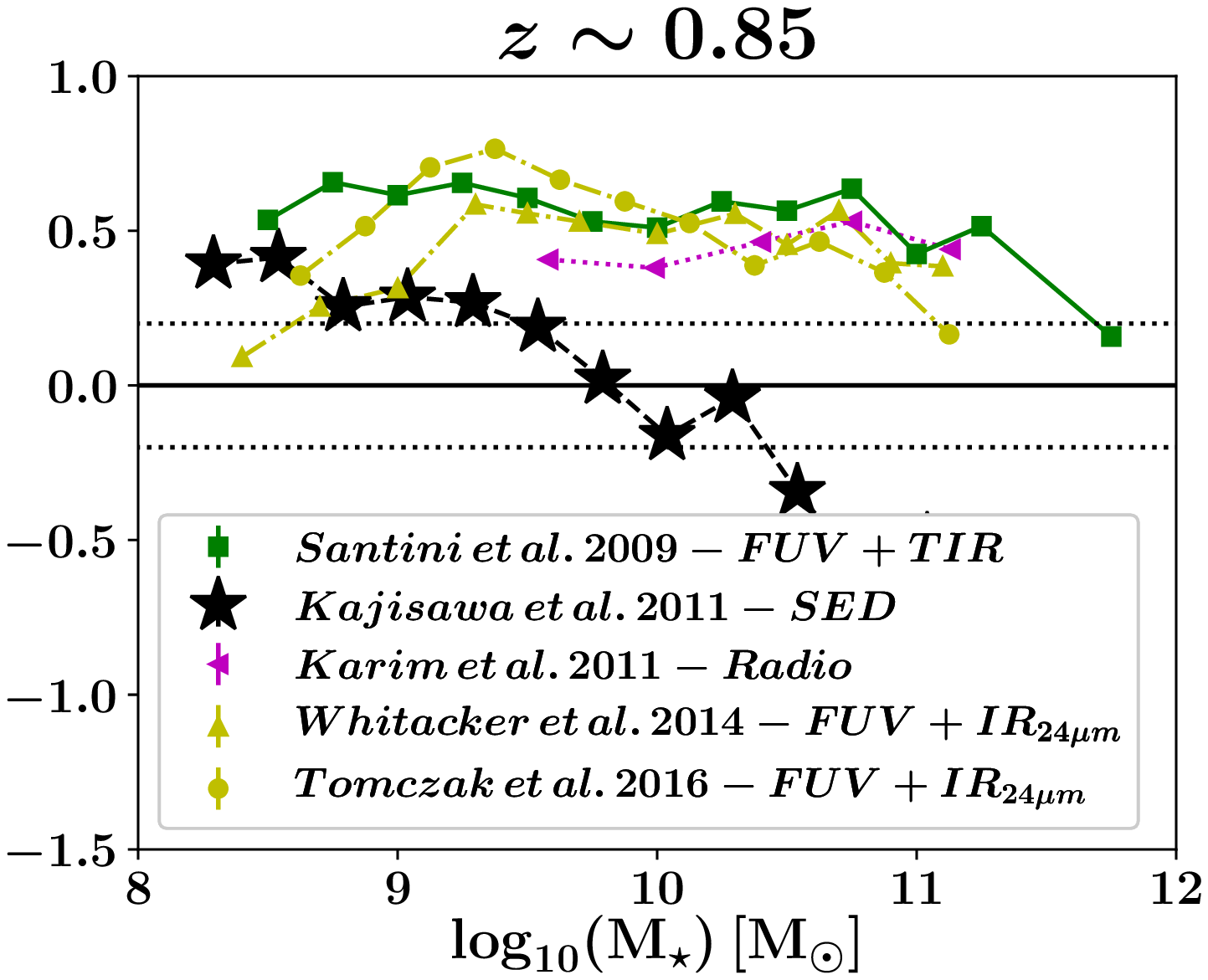}
\caption{The offset, in dex, between a range of observations with respect the ${\rm SFR-M_{\star}}$ relation from the EAGLE simulation reference model with different panels showing different redshifts, ranging from $z \simeq 0.85$ to $4$. The 0 dex line represents the EAGLE reference model. The observed stellar masses when necessary were altered into the \citet{chabrier03} IMF and the conversion laws between luminosities and observed SFRs were updated to the \citet{Kennicutt2012} relations. Top left panel: The blue right pointing triangles represent the observations of \citet[][UV + IRX-$ \beta $]{Bouwens2014}, the green squares \citet[][FUV+TIR]{Heinis2014}, the black stars the observations of \citet[][SED fitting]{Salmon2015}, the orange circles \citet[][FUV+IR]{Tomczak2016} and the red diamonds the results from \citet[][UV + IRX-$ \beta $]{Santini2017}. Middle top Panel: The Magenta left pointing triangles represent the results from \citet[][Radio]{Karim2011}, the black stars \citet[][SED fitting]{Bauer11}, the dark green squares \citet[][FUV+TIR]{Bauer11},  orange triangles represent \citet[][FUV+IR]{Whitaker2014,Tomczak2016}.  Right top panel: The yellow circles represent the results from \citet{Daddi2009}. Note that other observational studies present in this panel are described in the previous panels. Middle bottom Panel: The green squares represent the results from \citet[][FUV+TIR]{Santini2009} black stars the observations from \citet[][SED fitting]{Kajisawa2010}. Observational studies report results which can differ by 0.2-1.2 dex. The EAGLE reference model is usually more consistent with the results reported by authors who used SED fitting (black stars) to derive both SFRs and stellar masses \citep{Katsianis2015} but the offset, even from these observations can be up to 0.4 dex.}
\label{fig:SFRIR1}
\end{figure*}

\section{The comparison between observed and simulated SFR$-{\rm M}_{\star}$ relations}
\label{thecode}

\subsection{EAGLE vs observations}
\label{thecode1}

The Evolution and Assembly of GaLaxies  and  their  Environments  simulations \citep[EAGLE,][]{Schaye2015,Crain2015,McAlpine2016} are a well studied suite of cosmological hydrodynamical simulations with the reference model being able to produce galaxies with realistic SFRs and stellar masses. It broadly reproduces the observed star formation rate function of ${\rm z = 0-8}$ galaxies \citep{Katsianis2017}, the evolution of the stellar mass function \citep{Furlong2014} and the  scatter of the  ${\rm sSFR}$-${\rm M_{\star}}$ relation \citep{Matthee2018,Katsianis2019,Davies2019} at $z \simeq 0-4$. The reference simulation spans a 100 co-moving Mpc per side in a cubic, periodic volume. The initial conditions were generated using  the IC$_{-}$2LPT$_{-}$GEN code  \citep{Jenkins2010}.  EAGLE-REF tracks the evolution of baryonic gas, stars, non baryonic dark matter particles and massive black holes from z = 127 to z = 0. It includes various physical prescriptions like SNe feedback \citep{DVecchia2012,Katsianis2017}, AGN feedback \citep{springetal05,Rosas2016}, metal cooling \citep{wiersma09b} and star formation \citep{Schaye2008} assuming a \citet{chabrier03} IMF. It follows $2 \times 1504^3$ particles with an equal number of gas and dark matter elements with initial mass of dark matter particles $m_{D} = 9.7 \times 10^6 M_{\odot} $ and particle gas mass of $m_{g} = 1.8 \times 10^6 M_{\odot} $. The reference simulation produce the observed molecular hydrogen abundances \citep{Lagos2015}, supermassive black holes evolution \citep{Rosas2016}, angular momentum evolution \citep{Lagos2017} and quenching histories of cluster galaxies \citep{Pallero2019}. However, the simulation is unable to reproduce the observed SFR$-{\rm M}_{\star}$ relation especially at $z \simeq 1-2$ \citep{Furlong2014}. \citet{Katsianis2015}, demonstrated that the EAGLE, Illustris and ANGUS simulations alongside with semi-analytic models \citep{Dutton2010} produce almost identical relationships, indicating that the tension of simulations with observations is a common finding between different collaborations. The discrepancy between observed and simulated relations is typically -0.2 to 0.8 dex, depending on mass, redshift, sample selection method and observational technique used to derive SFRs and stellar masses, with the simulations predicting a factor of 2-4 smaller SFRs at a fixed ${\rm M}_{\star}$ than observed.

In Fig. \ref{fig:SFRIR1} we present the offset of a range of observations with respect to the EAGLE reference model (represented by the black 0 dex line). In the top left panel ($z \simeq 4.0$) the blue triangles represent the observations of \citet[][UV + IRX-$ \beta $]{bouwens2012}, the green squares \citet[][FUV + TIR]{Heinis2014}, the black stars the observations of \citet[][SED fitting]{Salmon2015}, the orange circles \citet[][FUV+IR]{Tomczak2016} and the red diamonds the results from \citet[][UV + IRX-$ \beta $]{Santini2017}. We note that in order to perform a consistent and up to date comparison between observational studies and EAGLE, the observed stellar masses when necessary were altered into the \citet{chabrier03} IMF and the conversion laws between luminosities and observed SFRs were updated to the \citet{Kennicutt2012} relations. We also note that the observed relations and the comparison between them does not change significantly after the above calibrations \citep{Katsianis2015}. We can see that the observations of \citet{Heinis2014} and \citet{bouwens2012} differ from the EAGLE reference model by $ \simeq 0.5-1$ dex. However, the \citet{Salmon2015} and \citet{Santini2017} observations are within $ \simeq 0-0.3$ dex from the predictions. This behavior is found at all redshifts with the reference EAGLE model and observations having offset star formation rates from $-0.2$ to $1.0$ dex depending on masses and redshifts. However, we note that there is a similar tension between the observed ${\rm SFR}$-${\rm M_{\star}}$ relation reported by different authors. For example, \citet{Heinis2014} and \citet{Salmon2015} results differ by 0.6-0.8 dex at $z \simeq 4$. Different authors use different diagnostics, assumptions and wavelengths to infer galaxy SFRs. Thus, it is interesting to derive ${\rm SFR}$-${\rm M_{\star}}$ relations using a set of artificial/simulated galaxies for which we have access to their SFRs, stellar masses and full spectral energy distributions. We can then mimic the methodologies used by different observational studies and explore further the inconsistency between hydrodynamic simulations and observations and the discrepancy between the results reported by a range groups.

We have to note that selection effects, besides the criteria used to define MS objects \citep{Renzini2015}, also can affect any comparison between observational studies \citep{Speagle14} and can enhance the disagreement with simulations \citep{Katsianis2015}. Some ``parent'' selection methods commonly used in the literature include the B-z vs z-K (sBzK) technique \citep{Daddi2004,Daddi2007,Kashino2013}, the Lyman break technique \citep{bouwens2012} and cuts on the color-magnitude diagram \citep{Elbaz2007}. The above methods pre-select star forming galaxies and steeper slopes are expected for the derived  ${\rm SFR}$-${\rm M_{\star}}$, since a large portion of less active galaxies that would be classified as star forming is prematurely excluded\footnote{\citet{Speagle14} pointed out that the normalization of the MS does not differ significantly between studies which use different parent selection methods. However, the logarithmic slope $\alpha$ differs by $\pm 0.5$ from study to study and is typically larger for pre-selected parent star forming objects. \citep{oliver2010,Sobral2011,Karim2011,Whitaker2012,Speagle14}, well before a MS is defined.} We choose to neglect the effect of parent sample selection in our comparisons with simulations, following previous studies \citep{Sparre2014,Furlong2014}. Complicating further our analysis by reckoning numerous sample selection criteria that are greatly different from study to study would divert our focus from the main goal of our work which is to investigate the impact of the employed methodology to derive galaxy properties using mock galaxies on the ${\rm SFR}$-${\rm M_{\star}}$ relation.

\section{The EAGLE+SKIRT data}
\label{thecode2}

\citet{Camps2018} performed full 3D radiative transfer postprocessing simulations applying the SKIRT code \citep{Baes2003,Baes2011,Camps2015} on the EAGLE galaxies. The authors calculated mock observables that fully took into account the absorption, scattering and thermal emission from the EAGLE simulation.  Bellow we briefly describe the procedure.

For each stellar particle, a SED was assigned which was acquired from the GALEXEV library \citep{Bruzualch03}, based on the mass of the particle, age and metallicity. For each star forming particle, a SED was acquired from the MAPPINGS III templates \citep{Groves2008} based on its SFR, pressure of the interstellar medium, compactness, covering fraction of the photo-dissociation region and metallicity. MAPPINGS models are used to describe the dusty $H_{II}$ regions. The dust distribution is obtained from the distribution of gas while the assumed model is \citet{zubko2004}. The dust mass is derived from the cool and star-forming gas, and correlates with the fraction of metals in dust ($f_{dust}$). The adopted values for the covering fraction, the dust-to-metal ratio and $f_{dust}$ are based on the following scaling relations: 1) the sub-mm colour diagram, 2) the specific dust mass ratio versus stellar mass and 3) the NUV-r colour relation. The calibration was done between galaxies from the Herschel Reference Survey \citep[HRS, ][]{Boselli2010,Cortese2012} and a matched sub-sample of 300 EAGLE galaxies \citep{Camps2016}. The adopted value of covering fraction is $f_{PDR}$ = 0.1. The metal fraction is set to be $f_{dust} = 0.3$ \citep{Brinchmann2013}. The dust density distribution of the system is discretised over an octree grid \citep{Saftly2013}. Physical quantities, such as the radiation field and dust density, are assumed to be constant. The smallest possible cell is 60 pc on a side. In order to perform the radiative transfer simulation it is important to have a sufficiently resolved dust distribution. Thus, the EAGLE+SKIRT sample excludes galaxies with low SFRs which have little or no dust \citep{Camps2018} \footnote{We note that the above pre-selection criteria could exclude some realistic objects but the offset between the ${\text{SFR}}$-${\rm M_{\star}}$ relations derived from the EAGLE+SKIRT data and the full EAGLE data is small ($ \simeq 0.05$ dex at $z = 4$, $  \simeq 0.08$ dex at $z = 2$ and $ \simeq 0.05$ dex at $z = 1$). Thus any comparison between the observed and EAGLE+SKIRT  ${\text{SFR}}$-${\rm M_{\star}}$ relations  at the  $\log_{10}(M_{\star}/M_{\odot}) \simeq 8.5-11.0$ range is not significantly affected by the selection criteria described in \citet{Camps2018}.}

The input SEDs and dust properties are sampled on a single wavelength grid that performs the radiative transfer calculations. Photon packages are given wavelengths which correspond to the grid points, dust absorption and re-emission. The output fluxes are recorded on the same grid which has 450 wavelenght points from 0.02 to 2000 $\mu$m on a logarithmic scale. The band-integrated fluxes and absolute magnitudes that were produced correspond to the following filters: GALEX FUV/NUV \citep{Morrisey2007}, SDSS$_{ugriz}$ \citep{Doi2010}, 2MASS JHK \citep{Cohen2003}, WISE W1/W2/W3/W4  \citep{Wright2010}, Spitzer MIPS  24/70/160 \citep{Rieke2004}, Herschel PACS  70/100/160 \citep{Poglitsch2010} and Herschel SPIRE  250/350/500 \citep{Griffin2010}. To obtain  the  integrated  fluxes, the  simulated SEDs were convolted with the instrument’s response curve. The procedure depends on whether the instrument counts photons or measures energy (bolometers) and is summarised in detail at the Appendix A of \citet{Camps2016}. To obtain broadband magnitudes in the rest frame the detected SEDs are convolted with the corresponding response curves while the resulting fluxes are converted to absolute AB magnitudes, taking into account the fixed assumed galaxy-detector distance of 20 Mpc (the median distance of the HRS sample). To obtain fluxes in the observer frame, the detected SEDs are redshifted and scaled following 
\begin{eqnarray}
\label{1}
 {\rm f_{v, obs} = (1+z) \left(\frac{20 \, Mpc}{DL} \right)^2 f_{v, shifted},}
\end{eqnarray}
where z is the galaxy’s redshift and $D_{L}$ the corresponding luminosity distance. The $D_{L}$ used are given by \citet{Adachi2012} following \citet{Baes2017}.

Thus, the mock galaxy SEDs consist of UV to submm flux densities and rest-frame luminosities  for  almost 0.5  million simulated  galaxies,  from $z = 0$ to $6$. The above data have already been used to investigate the cosmic spectral energy distribution \citep{Baes2019}, the relation between the hosts of merging compact objects to properties of galaxies like metallicities, SFRs, stellar masses and colours \citep{Artale2019}, the $\sigma_{sSFR}-{\rm M_{\star}}$ relation \citep{Katsianis2019}, the nature of sub-millimeter and high-SFR systems \citep{McAlpine2019} and galaxy number counts at 850 $ {\rm \mu m }$ \citep{Cowley2018}. We use the same data to study how typical SFR and $M_{\star}$ diagnostics affect the ${\text{SFR}}$-${\rm M_{\star}}$ relation and to make a fairer comparison with the observations by using the same methods to infer SFRs and stellar masses for the simulated galaxies. We stress that the EAGLE objects that were post-processed by SKIRT were galaxies with stellar masses $\log_{10}(M_{\star}/M_{\odot}) > 8.5$,  above the resolution limit of 100 gas particles and with sufficient dust content.

\subsection{Stellar masses and SFRs from the EAGLE+SKIRT data}
\label{thecode11}

To infer stellar masses from the EAGLE+SKIRT galaxies, we use the Fitting and Assessment of Synthetic Templates (FAST) code \citep{Kriek2009} to fit the mock SEDs, following a similar procedure as various  observational studies \citep{Gonzalez2012,Botticella2017,Aird2017}. Following the same procedure as in \citet{Katsianis2019} we use the \citet{Bruzualch03} stellar population synthesis models and assume an exponentially declining SFH [${\rm SFR = exp(-t/tau)}$] \citep{Fumagalli2016,Abdurrouf2019}, the Chabrier IMF \citep{chabrier03}, the \citet{Calzetti2000} dust attenuation law \citep{Cullen2018,Mclure2018} and a metallicity Z = 0.2 ${\rm Z_{\odot}}$ \citep{Chan2016,Mclure2018b}. We note that these assumptions are motivated by observational studies but not necessarily stand neither for the real/observed nor the EAGLE+SKIRT simulated galaxies (in table \ref{tab_stepsfrf3} we sumarize the SED fitting assumptions used by different authors). We employ numerous wavelengths filters like GALEX$_{FUV}$, GALEX$_{NUV}$, SDSS$_{u}$, SDSS$_{g}$, SDSS$_{r}$, SDSS$_{i}$, SDSS$_{z}$, TwoMass$_{J}$, TwoMass$_{H}$, TwoMass$_{Ks}$, UKIDDS$_Z$, UKIDDS$_Y$, UKIDDS$_J$, UKIDDS$_H$, UKIDDS$_K$, Johnson$_U$, Johnson$_B$, Johnson$_V$, Johnson$_R$, Johnson$_I$, Johnson$_J$, Johnson$_M$,  WISE$_{W1}$, WISE$_{W2}$, WISE$_{W3}$, WISE$_{W4}$, IRAS$_{12}$, IRAS$_{25}$, IRAS$_{60}$, IRAS$_{100}$, IRAC$_{I1}$, IRAC$_{I2}$, IRAC$_{I3}$, IRAC$_{I4}$, MIPS$_{24}$, MIPS$_{70}$, MIPS$_{160}$, PACS$_{70}$, PACS$_{100}$, PACS$_{160}$, SPIRE$_{250}$, SPIRE$_{350}$ and SPIRE$_{500}$ in order to limit parameter degeneracies to the SED fitting procedure \citep{Katsianis2015,Santini2017}. 

To derive SFRs from the EAGLE+SKIRT data, we follow a range of techniques:
\begin{itemize}

\item 1) Employing the SED fitting technique in which the same bands used to derive the stellar masses are exploited \citep{Kriek2009}. We label the above as SFR${\rm_{SED-FAST}}$.

\item 2) Combining the TIR obtained from the $24 {\rm \mu m}$ luminosities and dust uncorrected FUV (1600 ${\rm \AA}$). The TIRs are obtained adopting the luminosity-independent conversion from $IR_{24 \mu m}$ \citep{Wuyts2008} following \citet{Franx2008}, \citet{Muzzin2010}, \citet{Whitaker2014} and \citet{Tomczak2016}. We convert the TIR luminosities and UV luminosities into SFRs following \citet{Kennicutt2012} \footnote{{$  Log_{10} (SFR_{TIR}) = {\rm Log_{10} (L_{TIR}) - 43.41}$ \\ $ Log_{10} (SFR_{FUV}) = {\rm Log_{10} (L_{\rm FUV}) - 43.35}$ }} while the total SFR is given by:
\begin{eqnarray}
  \label{eq_A4}
  \begin{matrix}
    {\rm SFR_{24 \mu m} } = {\rm SFR_{UV-uncor}} + {\rm SFR_{TIR_{24 \mu m}}}.
   \end{matrix}
\end{eqnarray}
We label the above as SFR${\rm_{24 \mu m - Wuyts \, et \, al. \, 2008}}$ $\,$.

\item 3) Combining the Total IR (TIR) luminosities with dust-uncorrected UV emission (1600 ${\rm \AA}$). The TIR luminosities are estimated from the 24, 70 and 160 ${\rm \mu m}$ MIPS luminosities following \citet{Verley2010} and \citet{Espada2019} and employing the relation given by the \citet{DaleHelou2002} templates \footnote{The coefficients of the $L_{TIR} = a L_{24 \mu m} + b L_{70 \mu m} + c L_{160 \mu m}$ relation were derived from a singular value decomposition solution to an overdetermined set of linear equations. The equation matches the model bolometric infrared luminosities, for all model SED shapes, from 1-4$\%$ at $z = 0-4$.}).
We convert the TIR and dust uncorrected FUV luminosities into SFRs  using \citet{Kennicutt2012} while the total SFR is obtained from: 
\begin{eqnarray}
  \label{eq_A3}
  \begin{matrix}
  {\rm SFR_{24, 70, 160 \mu m}} = {\rm SFR_{UV-uncor}} + {\rm SFR_{TIR_{24, 70, 160 \mu m}}}.
   \end{matrix}
\end{eqnarray}
We label the above as SFR${\rm _{24, 70, 160 \mu m - r \, Dale \& Helou \, 2002}}$.

\item 4) Using the luminosity emitted by dust derived from the 250, 350 and 500 ${\rm \mu m}$ fluxes, the code CIGALE \citep{Boquien2019} and the \citet{Dale2014} templates combined with the uncorrected FUV light. The dust luminosities and UV luminosities were converted to SFRs using the \citet{Kennicutt2012} relations. In a similar framework  \citet{Heinis2014} inferred the dust luminosities of the COSMOS galaxies by adjusting the 250, 350 and 500 ${\rm \mu m}$ fluxes to the \citet{DaleHelou2002} templates, using an older version of CIGALE \citep{Noll2009} and the \citet{kennicutt1998} relations \footnote{The  dust  templates of \citet{Dale2014} are based on the same  sample of nearby star–forming galaxies originally presented in \citet{DaleHelou2002}}. The authors combined the above with FUV luminosities (1570$-$1620 ${\rm \AA}$) in order to derive the galaxy SFRs. We label the above as SFR${\rm _{250, 350, 500 \mu m - C \, Dale \& Helou \, 2014}}$.

\item 5) Employing the FUV luminosities (e.g. 1600 ${\rm \AA}$) dust-corrected using the IRX-${\rm \beta}$ relation \citep{meurer1999}. In order to obtain the FUV SFRs we follow the method described in \citet{smit12} and \citet{Katsianis2016}. We correct the FUV luminosities assuming the infrared excess (IRX)-$\beta$ relation of \citet{meurer1999}:
\begin{eqnarray}
  {\rm A_{\rm 1600} = 4.43 + 1.99\,\beta,}
  \label{eq_A16}
\end{eqnarray}
where $A_{\rm 1600}$ is the dust absorption at 1600 $\AA$ and $\beta$ is the UV-continuum spectral slope. We assume a linear relation between $\beta$ and the luminosity \citep{bouwens2012,Tacchella2013}:
\begin{eqnarray}
  \langle\beta\rangle=\frac{{\rm d}\beta}{{\rm d}M_{\rm UV}}\left(M_{\rm
      UV,AB}+19.5\right)+\beta_{M_{\rm UV}},
  \label{eq_beta}
\end{eqnarray}

We assume the same ${\rm \left < \beta \right >}$ as \citet{Arnouts2005,Oesch2010,smit12,Tacchella2013,Katsianis2016} and \citet{Katsianis2017}\footnote{$ \beta  =-0.11(M_{UV,AB}+ 19.5)-2.00$ at $z \simeq 4.0$ \\ $ \beta  =-0.13(M_{UV,AB}+ 19.5)-1.70$ at $z \simeq 2.0$ \\ $\beta  =-0.13(M_{UV,AB}+ 19.5)-1.55$ at $z \simeq 1.0$}. Then, following \citet{HaoKen} we assume
\begin{eqnarray}
\label{eq_A17}
{\rm L_{\rm UV-{uncor}}} = {\rm L_{\rm UV_{corr}}e^{-\tau_{UV}}},
\end{eqnarray} 
where ${\rm \tau_{UV}}$ is the effective optical depth (${\rm \tau_{UV}}={\rm A_{\rm 1600}/1.086}$).
We convert the dust-corrected UV luminosities into SFRs following \citet{Kennicutt2012}
\begin{eqnarray}
\label{eq_A19}
{\rm Log_{10} (SFR) } = {\rm Log_{10} (L_{\rm UV_{corr}}) - 43.35}.
\end{eqnarray}
We label the above as SFR${\rm_{UV+IRX-\beta}}$.

\end{itemize}
All the above methods have been commonly used in the literature to derive SFRs but have different limitations. UV provides a direct measure of SFR, but could underestimate the total SFR due to dust attenuation effects \citep{Dunlop2017}. IR wavelengths (especially Mid-IR and Far-IR) are used to determine the total IR luminosity (TIR), which is used to trace star formation. A major drawback of IR studies is that they usually do not have sufficient wavelength coverage especially at FIR wavelengths \citep{Lee2013,Pearson2018}. In order to overcome this limitation to determine the TIR luminosities, other authors have relied on extrapolations from the available wavebands  \citep[e.g. Spitzer 24 ${\rm \mu m}$, ][]{Wuyts2008}. However, the 24 ${\rm \mu m}$ band, Mid-IR and Far-IR lumininosities can be compromised by AGN \citep{Brand2006,Ichikawa2012,Roebuck2016,Brown2019}. Even studies which have access to a range of IR wavelengths still have to rely on SED libraries \citep{DaleHelou2002}, which have been constructed  from  galaxies  at  low redshifts. These templates/models may not be representative for high-redshift objects. One other disadvantage of using TIR as a SFR tracer is that other sources can contribute to the heating of dust in galaxies and this contribution can be falsely interpreted as star formation. In particular, old stellar populations can significantly contribute to dust heating, complicating the relation between SFR and TIR emission \citep{Bendo2010,Boquien2011,Bendo2012,Viaene2017,Nersesian2019}. Due to the above limitations in the infrared other studies use SED fitting to bands beyond IR including UV wavelengths \citep{Leja2019,Hunt2019}. However, \citet{Santini2017} suggested that this method suffers from parameter degeneracies, which are serious for the SFR determination, and instead used dust-corrected UV luminosities in their analysis.

\begin{table*}
\centering
\resizebox{0.95\textwidth}{!}{%
  \begin{tabular}{ccccc}
    \hline \\
    & {\large Authors / Parent sample selection}  &
    {\Large ${\rm SFR}$} & {\Large ${\rm M_{\star}}$}  \\ & \large -main sequence selection  \\
    \hline \hline
    &  & Observations  & \\
    \hline \hline
    & \citet{Santini2009} / Optical-2$\sigma$ &  $2700 {\rm \AA}$ +  IR$_{24 \mu m}$, \citet{DaleHelou2002} &  SED, \citet{Bruzualch03} \\
    & Bruzual $\&$ Charlot (2003) models, & \citet{salpeter55} IMF, exponentially declining SFHs & dust extinction \citet{Calzetti2000}, 1 ${\rm Z_{\odot}}$\\
    \hline
    & \citet{Kajisawa2010} / K band-${\rm M_{\star}}$  &  $2800 {\rm \AA}$ + SED dust Correction & SED,  GALAXEV \citep{Bruzualch03} \\
    & Bruzual $\&$ Charlot (2003) & \citet{salpeter55} IMF, exponentially declining SFHs & dust extinction \citet{Calzetti2000}, 0.02-1 ${\rm Z_{\odot}}$ \\
    \hline
    & \citet{Bauer11} / H band-${\rm M_{\star}}$ & $2800 \AA $ + SED \citet{Calzetti2000} & SED,  HYPERZ \citet{Bolzonella2010} \\
    & Bruzual $\&$ Charlot (2003) & \citet{salpeter55} IMF, exponentially declining SFHs & dust extinction \citet{Calzetti2000}, 0.0001-0.05 ${\rm Z_{\odot}}$ \\
    \hline
    & \citet{Heinis2014} / i band-UV& $1600 {\rm \AA} $ + 250, 350, 500 ${\rm \mu m}$, \citet{DaleHelou2002} & SED, CIGALE\\
    & Bruzual $\&$ Charlot (2003) & \citet{chabrier03} IMF, exponentially declining SFHs  & dust extinction \citet{meurer1999} \\
    \hline
    & \citet{Steinhartdt2014} / UV-${\rm M_{\star}}$ & FIR, \citet{Casey2012} & SED, LePHARE \citep{Arnouts2011}  \\
    & Bruzual $\&$ Charlot (2003) & \citet{chabrier03} IMF, exponentially declining SFHs & dust extinction \citet{Calzetti2000}, 0.5 ${\rm Z_{\odot}}$\\
    \hline
    & \citet{Whitaker2014} / IR-UVJ & $2800 {\rm \AA} $ + IR$_{24 \mu m}$,\citet{Wuyts2008} & SED, FAST \\
    & Bruzual $\&$ Charlot (2003) & \citet{chabrier03} IMF, rising+declining exponantially SFHs & dust extinction \citet{Charlot2000},  1 ${\rm Z_{\odot}}$ \\
    \hline
    & \citet{Salmon2015} / photometric-${\rm M_{\star}}$ & Bayesian SED fitting  & Bayesian SED fitting \\
    & \citet{Bruzual2011}, & \citet{salpeter55} IMF, constant SFHs &  dust extinction \citet{Charlot2000}, 0.2 ${\rm Z_{\odot}}$\\
    \hline
    &  \citet{Tomczak2016} / K band-UVJ & $2800 {\rm \AA} $ + IR$_{0.3-8 \mu m}$, \citet{Wuyts2008} & SED, FAST \\
    & Bruzual $\&$ Charlot (2003) & \citet{chabrier03} IMF, exponentially declining SFHs & dust extinction \citet{Calzetti2000}, 1 ${\rm Z_{\odot}}$\\
    \hline
    &  \citet{Santini2017} H band-2$\sigma$  & $1600 {\rm \AA}$ + IRX-$\beta$, \citet{meurer1999} & SED, N/A \\
    & Bruzual $\&$ Charlot (2003) & \citet{salpeter55} IMF, rising+declining delayed SFHs & dust extinction \citet{Calzetti2000}, 0.02 ${\rm Z_{\odot}}$ \\
    \hline
    & \citet{Pearson2018} K band-Gaussian & SED, CIGALE & SED, CIGALE\\
    & Bruzual $\&$ Charlot (2003) & \citet{chabrier03} IMF, exponantially delayed declining SFHs & dust extinction \citet{Charlot2000},  0.02 ${\rm Z_{\odot}}$\\
    \hline \hline
    &  & EAGLE+SKIRT  & \\
    \hline \hline
    & Fig. \ref{fig:SFRSED}   & SED, FAST & SED, FAST \\
    \hline
    & Left panels of Fig. \ref{fig:SFRIR}   & $1600 {\rm \AA}$ + IR$_{24 \mu m}$, \citet{Wuyts2008} & SED, FAST \\
    \hline
    & Middle panels of Fig. \ref{fig:SFRIR}   & $1600 {\rm \AA}$ + 24, 70, 160 $ {\rm \mu m} $, \citet{DaleHelou2002} & SED, FAST \\
    \hline
    & Right panels of Fig. \ref{fig:SFRIR}   & $1600 {\rm \AA}$ + 250, 350, 500 $ {\rm \mu m} $, \citet{Dale2014} & SED, FAST \\
    \hline
    & Fig. \ref{fig:SFRUV}   & $1600 {\rm \AA}$ + IRX-$\beta$,\citet{meurer1999} & SED, FAST\\
    \hline
    & Figs. \ref{fig:SFRSED}, \ref{fig:SFRIR} and \ref{fig:SFRUV} (dotted line)   & SFR$_{Int}$ & $M_{\star, Int}$\\
\hline \hline
  \end{tabular}%
}
\caption{The methodologies used to infer SFRs and stellar masses in the compilation of observations and EAGLE+SKIRT data used in this work. Stellar masses are typically inferred by the SED fitting technique, which employs various assumptions. In this work we employ the \citet{Bruzualch03} models and assume an exponentially declining SFH [${\rm SFR = exp(-t/tau)}$] \citep{Fumagalli2016,Abdurrouf2019}, the Chabrier IMF \citep{chabrier03} with cutoffs at 0.1 and 100 ${\rm M_{\odot}}$ , the \citet{Calzetti2000} dust attenuation law \citep{Cullen2018,Mclure2018} and a metallicity of 0.2 ${\rm Z_{\odot}}$ \citep{Chan2016,Mclure2018b}. These choices are typical among the observational studies used in this work. When necessary we convert the IMFs of the observed relations from \citet{salpeter55} IMF to \citet{chabrier03} by decreasing the observed stellar masses by 0.21 dex \citep{Dave08,Santini12,Madau2014,Katsianis2015} while SFR conversion laws are re-calibrated to \citet{Kennicutt2012}.}
  \label{tab_stepsfrf3}
\end{table*}

\begin{figure}
\vspace{1.5 cm}
\includegraphics[scale=0.45]{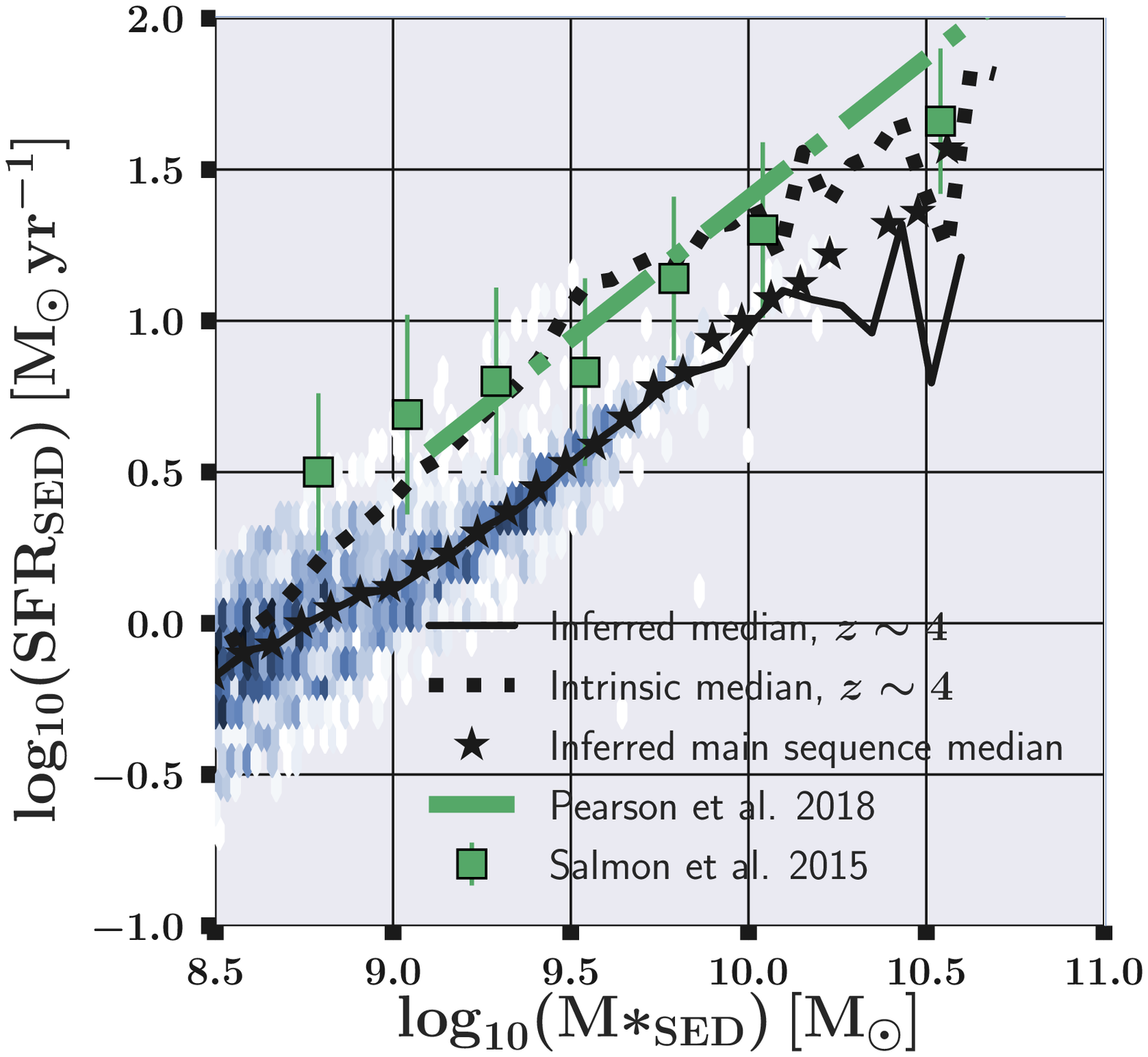} \\
\vspace{0.8 cm} \\
\includegraphics[scale=0.45]{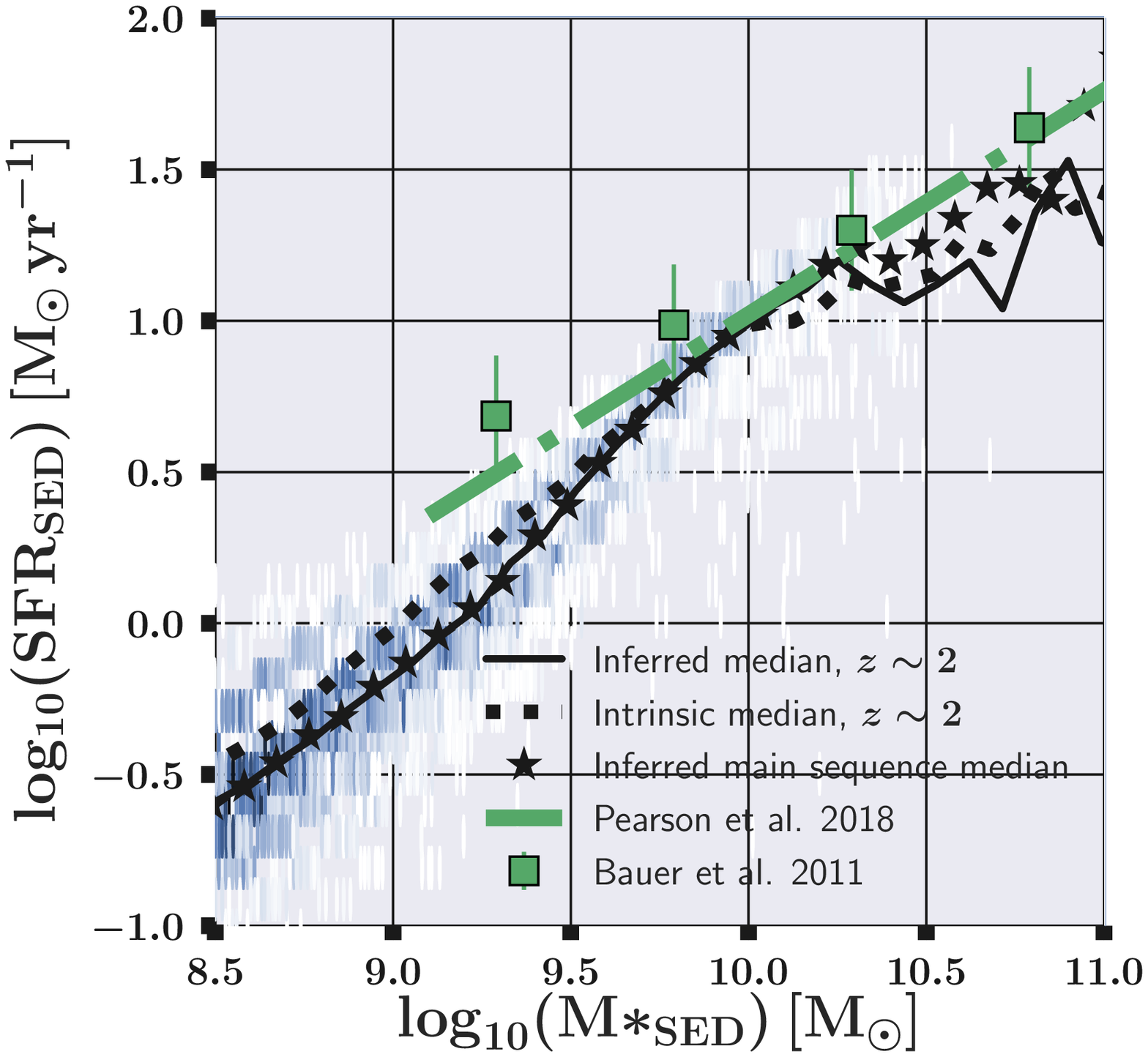} \\
\vspace{0.8 cm} \\
\includegraphics[scale=0.45]{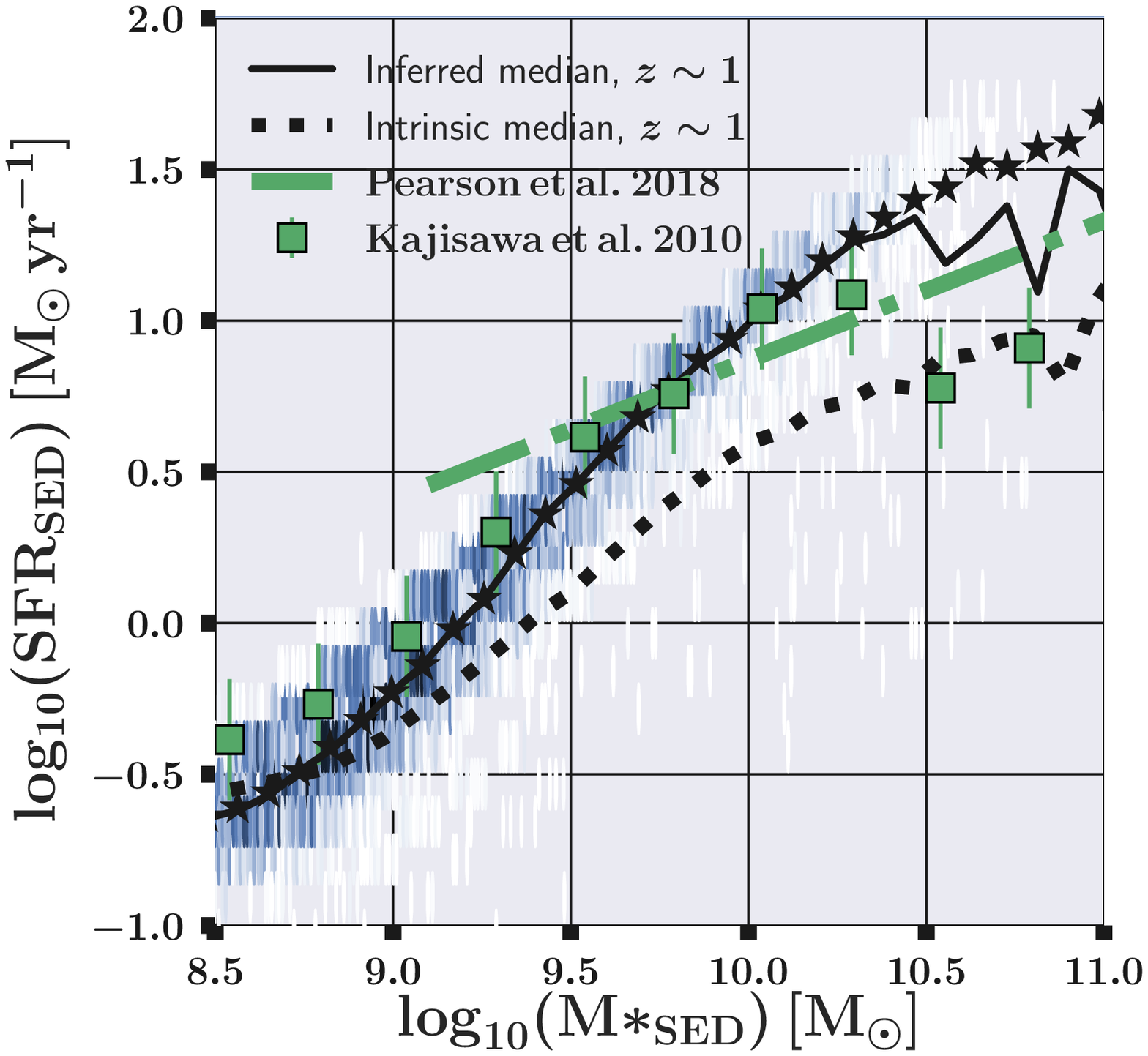}
\caption{ Black solid curves show the median ${\rm SFR-M_{\star}}$ relation using SED fitting  \citep{Kriek2009} to infer SFRs and stellar masses. The dotted line represents the intrinsic relation for the same galaxies (${\rm SFR_{Intr}-M_{\star, Intr}}$). The black stars represent the inferred Main-sequence relation defined by the exclusion of passive objects with $sSFR <  10^{-9.1}$ at $z \sim 4$, $sSFR <  10^{-9.6}$ at $z \sim 2$ and $sSFR <  10^{-10.1}$ at $z \sim 1$.}
\label{fig:SFRSED}
\end{figure}

\begin{figure*}
\centering
\hspace{-0.20 cm}
\includegraphics[scale=0.35]{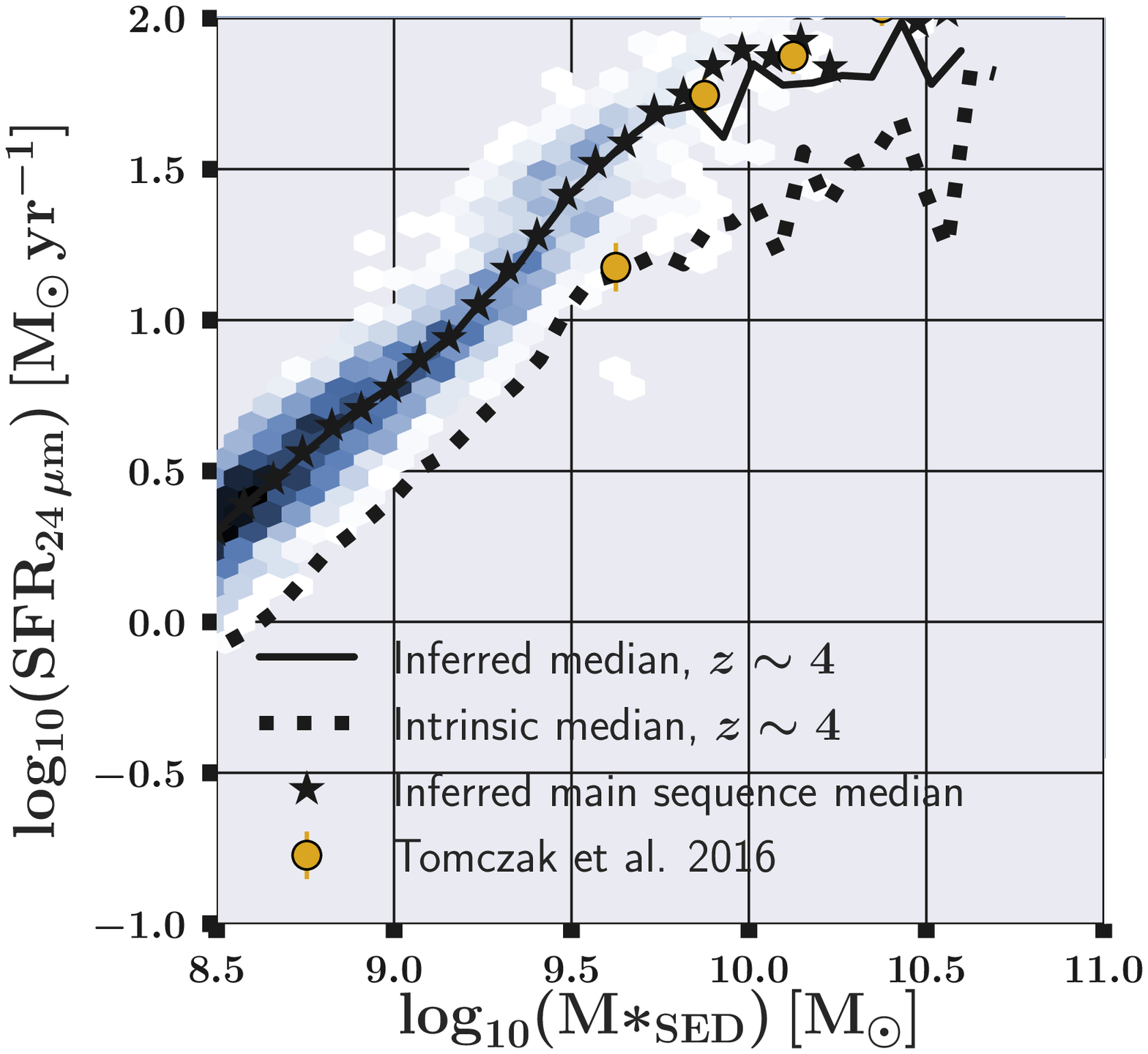}
\hspace{0.20 cm}
\includegraphics[scale=0.35]{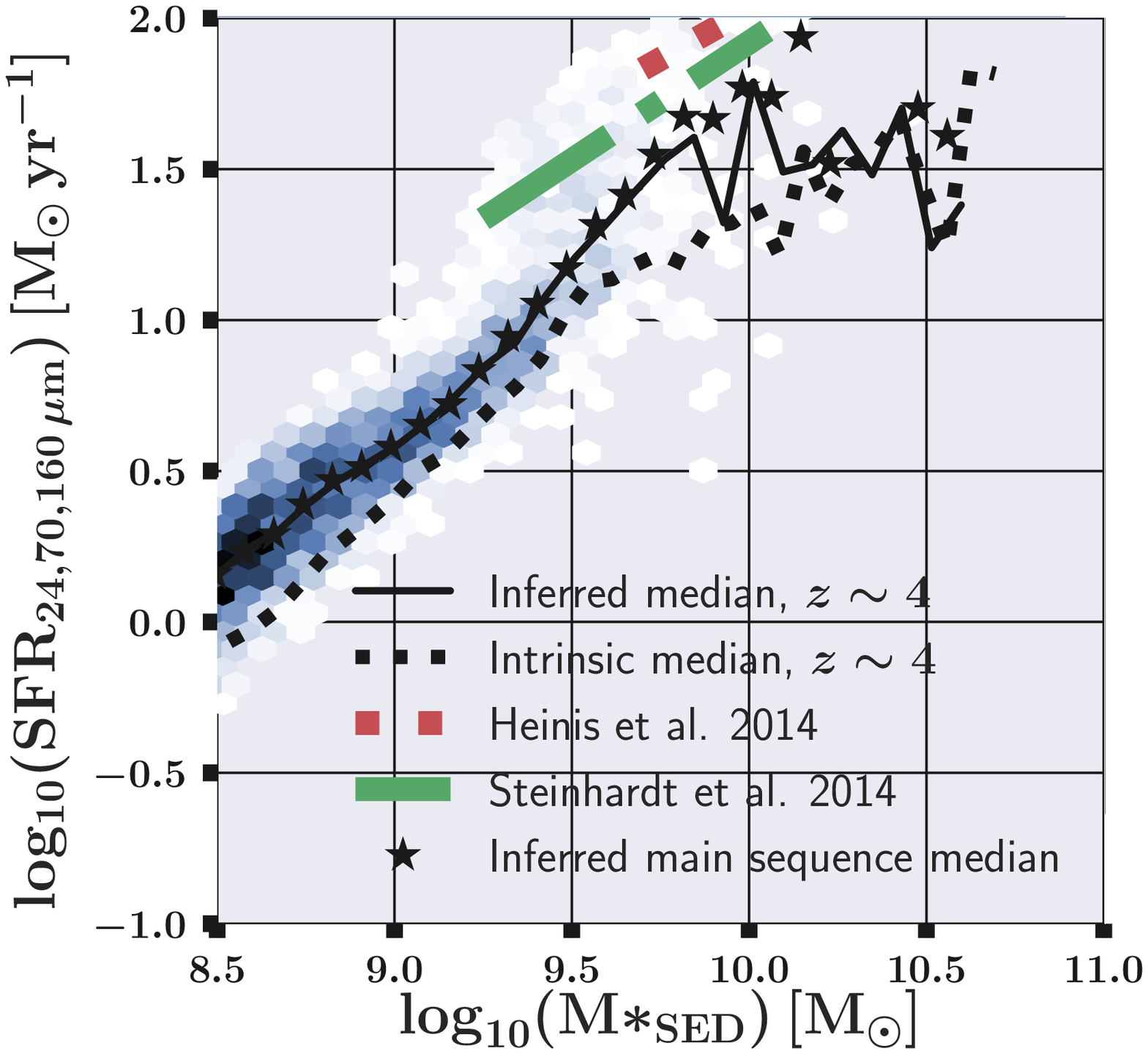}
\hspace{0.20 cm}
\includegraphics[scale=0.35]{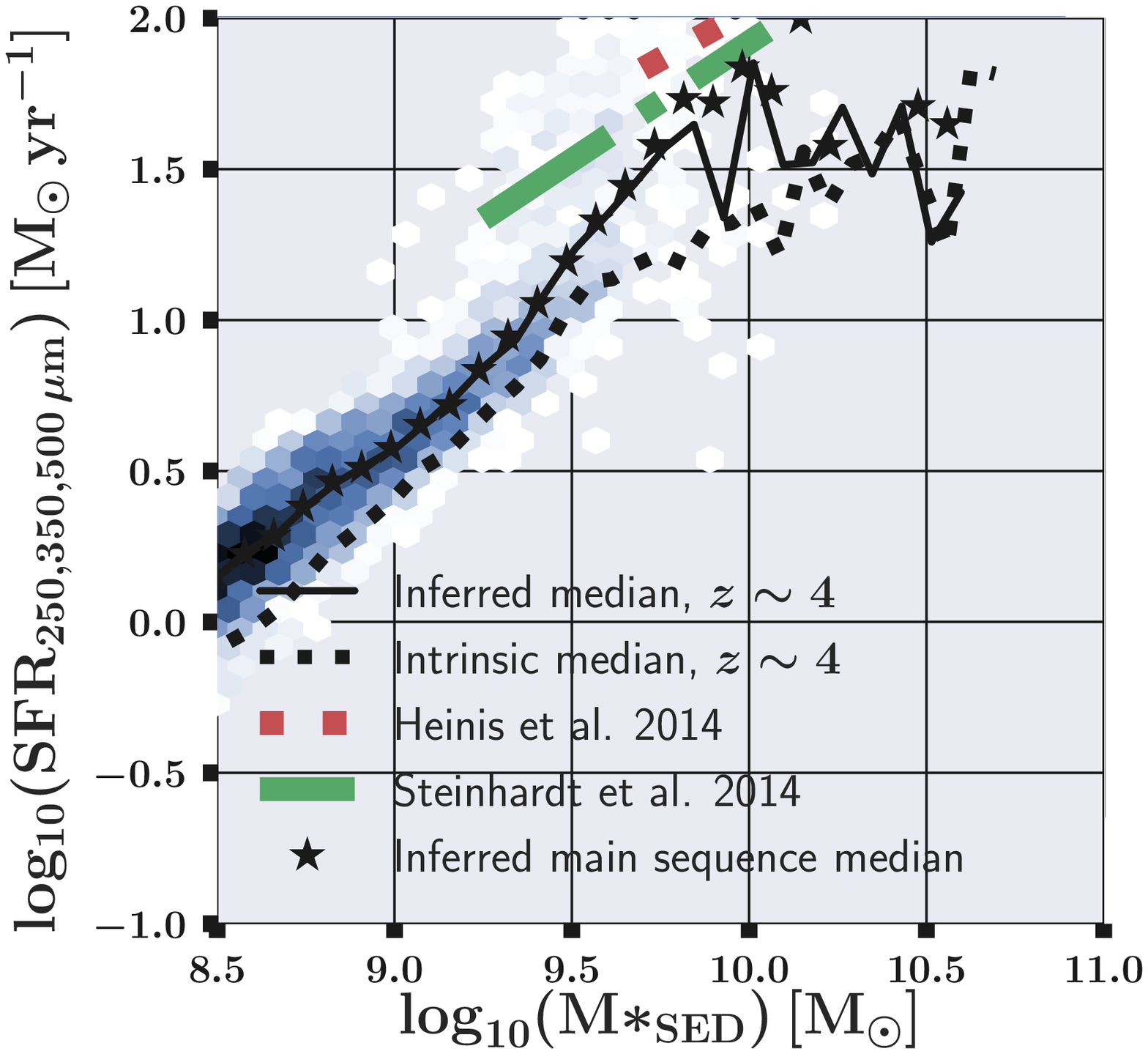} \\
\vspace{0.8 cm}
\includegraphics[scale=0.35]{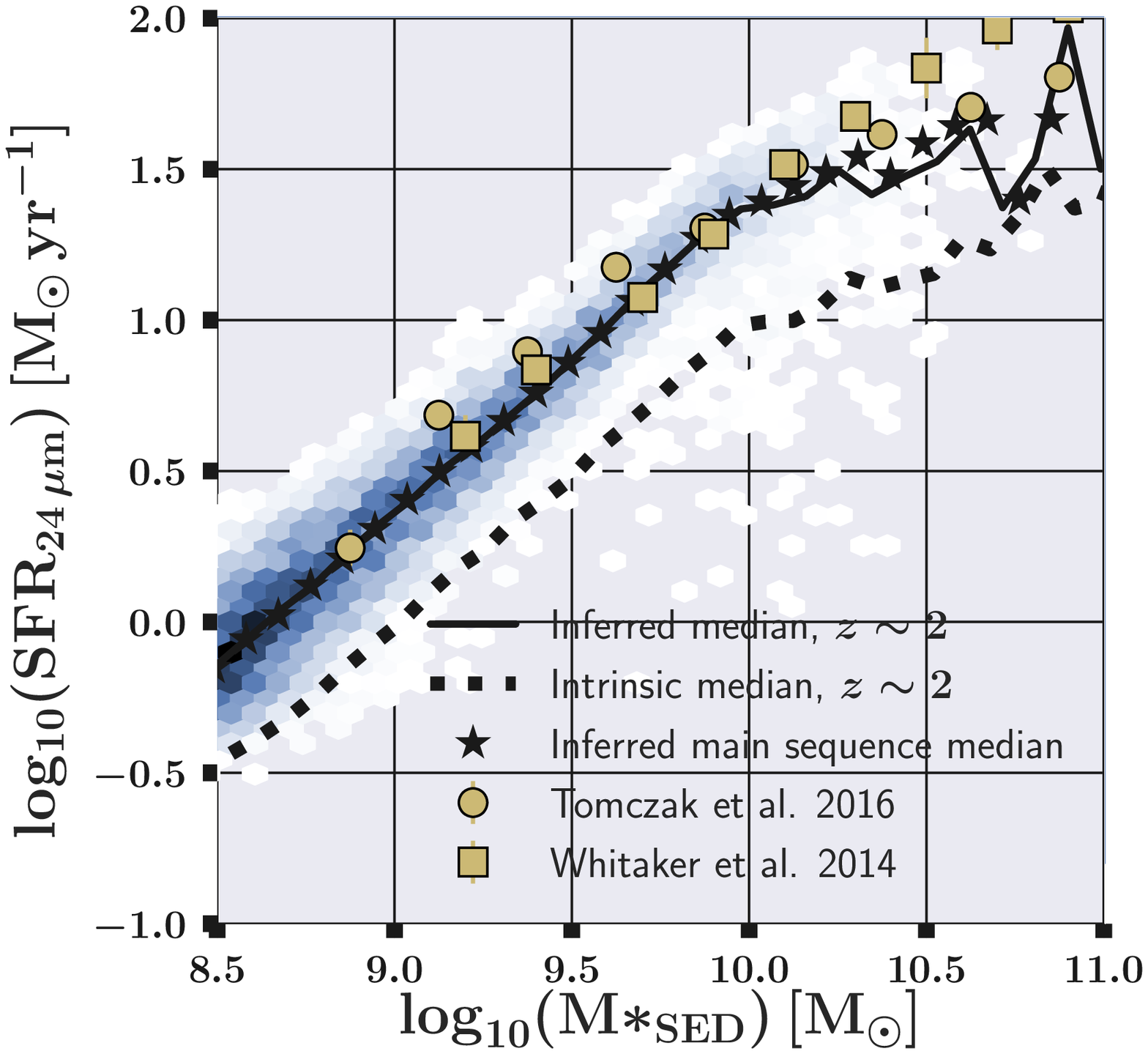}
\hspace{0.20 cm}
\includegraphics[scale=0.35]{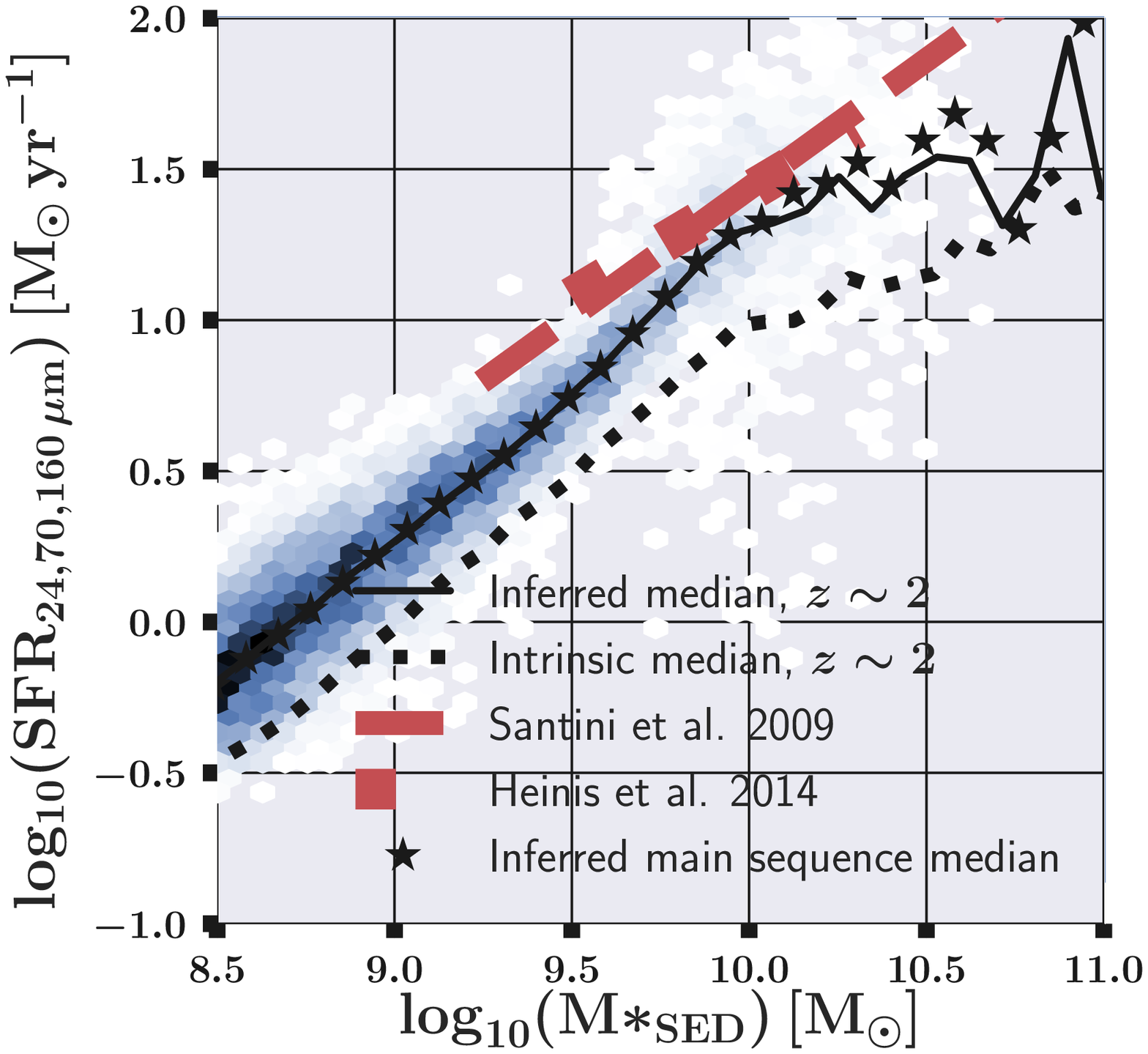}
\hspace{0.20 cm}
\includegraphics[scale=0.35]{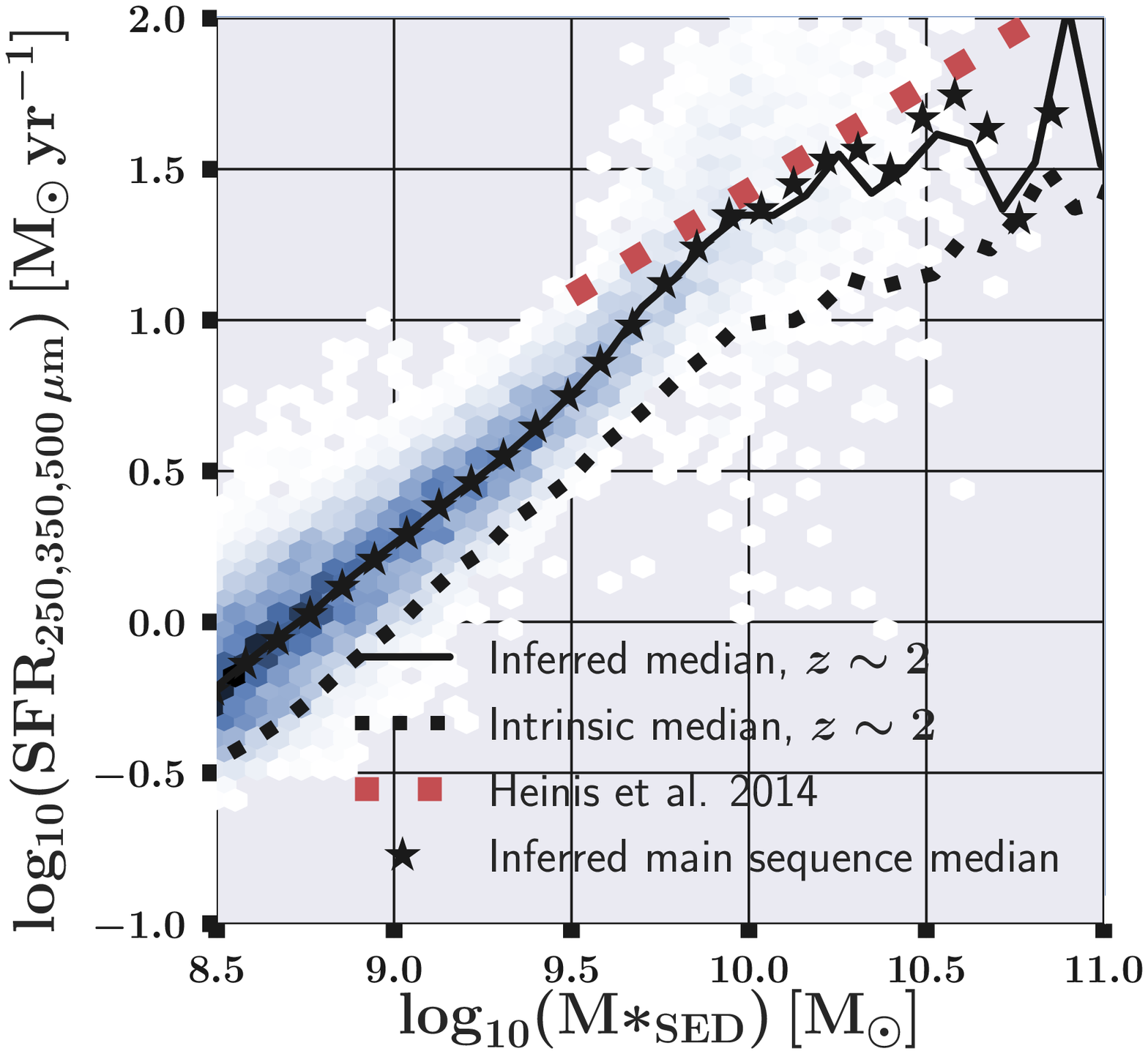} \\
\vspace{0.8 cm}
\includegraphics[scale=0.35]{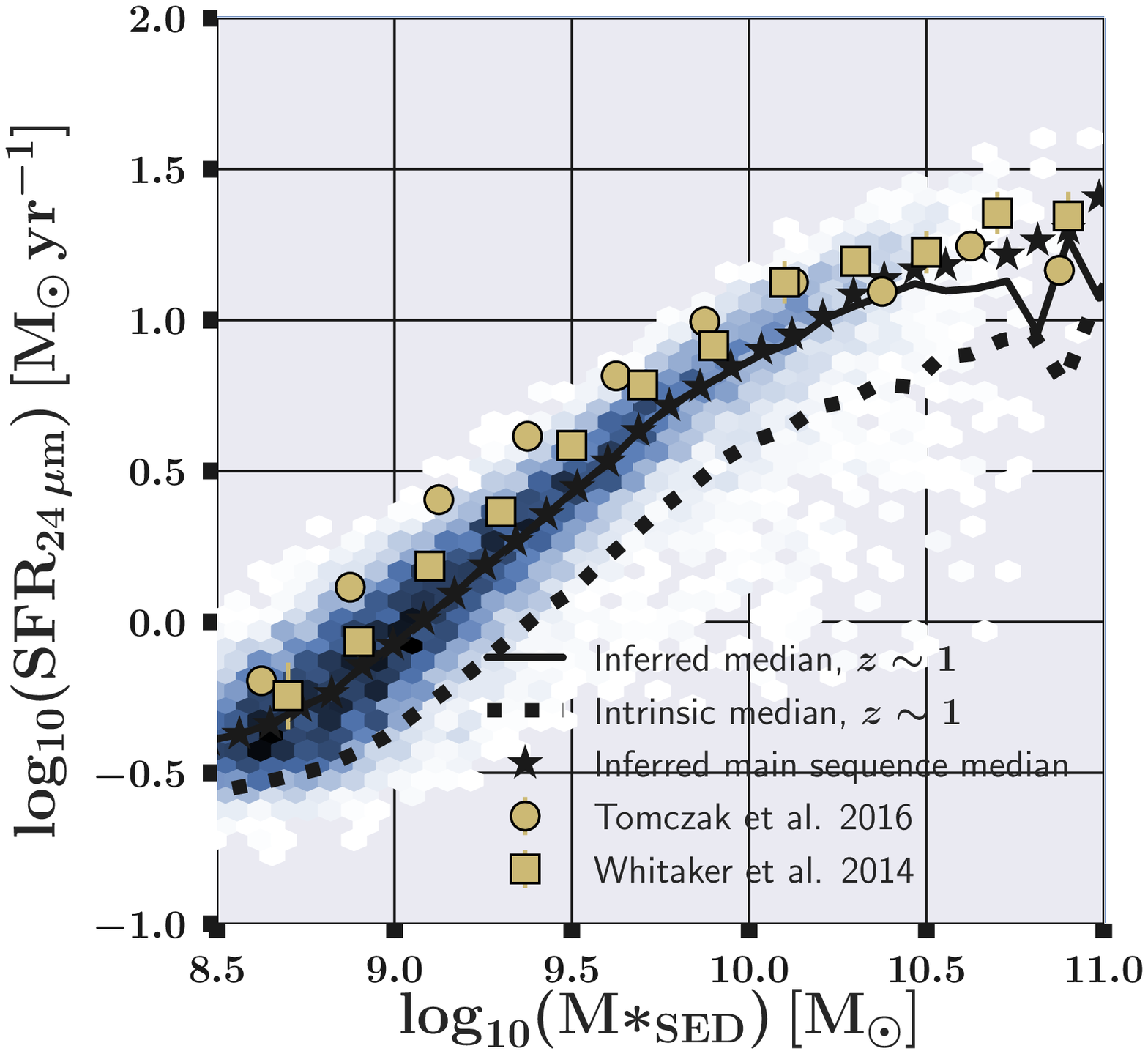}
\hspace{0.20 cm}
\includegraphics[scale=0.35]{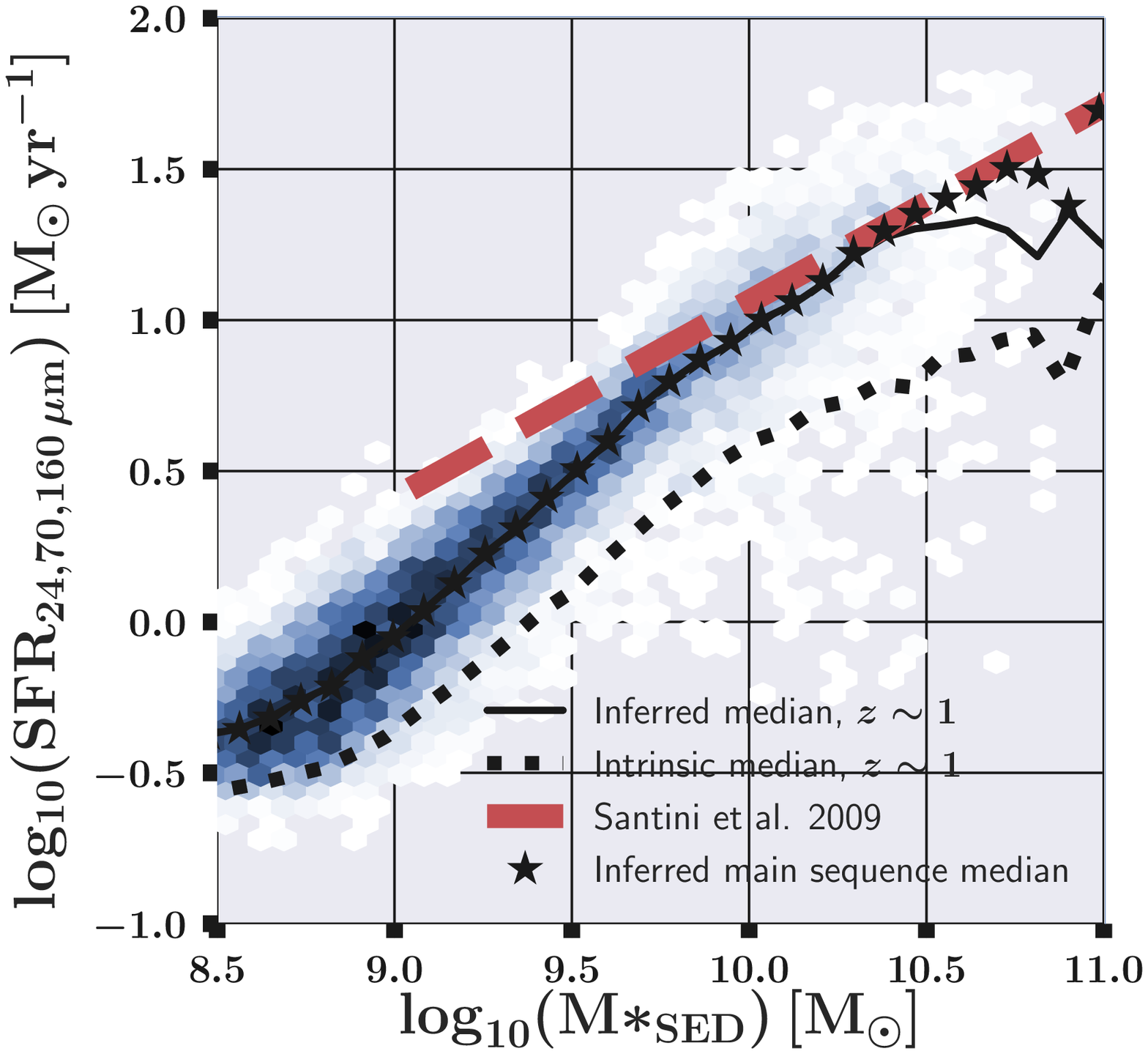}
\hspace{0.20 cm}
\includegraphics[scale=0.35]{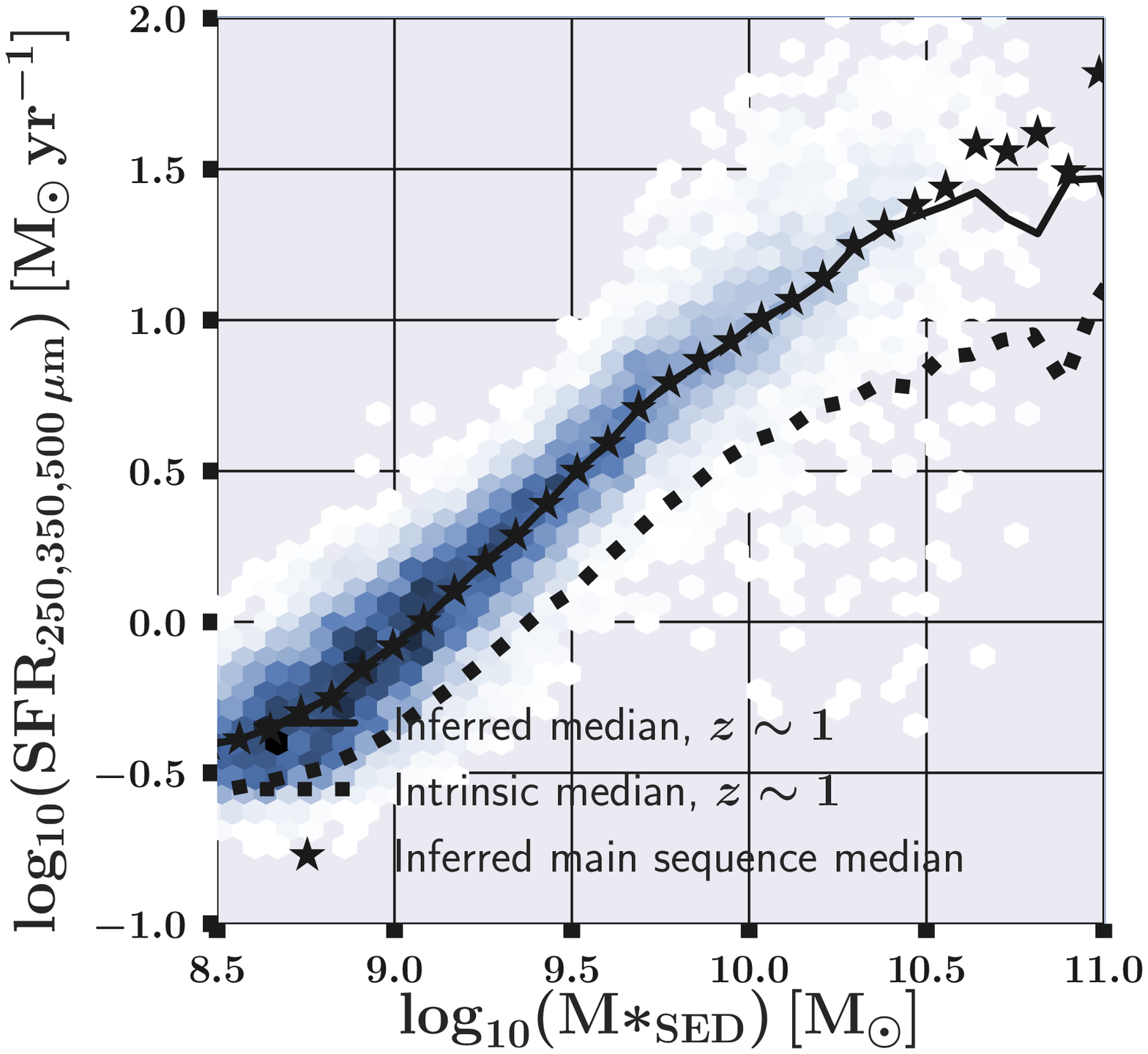}
\caption{The evolution of the ${\rm SFR-M_{\star}}$ relation using the EAGLE+SKIRT data using IR wavelengths. Black solid and dotted curves show the median relation inferred from the mock EAGLE+SKIRT observations, while the black dotted line represents the intrinsic relation (${\rm SFR_{Intr}-M_{\star, Intr}}$) for the same galaxies. The color scale indicates the number density of the EAGLE+SKIRT galaxies in the  ${\rm SFR-M_{\star}}$  plane. Different rows show different redshifts. Left panels: SFRs are calculated adopting the luminosity-independent conversion from the observed Spitzer/MIPS 24 ${\rm \mu m}$ flux density to the total IR luminosity following \citet{Wuyts2008}.  Stellar masses are calculated using the Fitting and Assessment of Synthetic Templates (FAST) code \citep{Kriek2009}. Middle panels: Star formation rates are calculated using the 24, 70 and 160 ${\rm \mu m}$ luminosities and their relation with the total IR luminosity given by the \citet{DaleHelou2002} templates and the TIR-SFR conversion given by \citet{Kennicutt2012}. Right panels: Star formation rates are calculated using the 250, 350 and 500 ${\rm \mu m}$ luminosities, the \citet{Dale2014} templates and the conversion given by \citet{Kennicutt2012}. The tension between observed and simulated  ${\rm SFR-M_{\star}}$ relations is generally highly reduced if both SFR and stellar masses are retrieved using similar methods in observations and simulations. In Table \ref{tab_stepsfrf4} we summarise the offset between the intrinsic and inferred relations at different mass bins. The black stars represent the inferred Main-sequence relation defined by the exclusion of passive objects with $sSFR <  10^{-9.1}$ at $z \sim 4$, $sSFR <  10^{-9.6}$ at $z \sim 2$ and $sSFR <  10^{-10.1}$ at $z \sim 1$.}
\label{fig:SFRIR}
\end{figure*}

\section{EAGLE+SKIRT vs observations}
\label{thecode3}

For the EAGLE+SKIRT galaxies in this work we investigate all the above methods. The compilation of observations and different techninques used in this work are described in Table \ref{tab_stepsfrf3}, while the results are summarized in Figs \ref{fig:SFRSED}, \ref{fig:SFRIR} and \ref{fig:SFRUV}, where we provide the number density plots of the inferred ${\text{SFR}}$-${\rm M_{\star}}$ plane and a comparison with observations (the density of points increases from white to dark blue). We note that the observations present at each panel alongside with the simulated results are derived following similar methods and wavelengths (table \ref{tab_stepsfrf3}). However, sample selection effects or unique assumptions for the SED modeling can be different from study to study and exploring these variations is beyond the scope of our current work.

\begin{itemize}

\item The black solid lines in the panels in Fig. \ref{fig:SFRSED} represent the median SFR${\rm_{SED-FAST}}$ $\,$ - $\,$ M${\rm_{\star, SED-FAST}}$ relation at $z \simeq 4$ (top), $z \simeq 2$ (middle) and $z \simeq 1$ (bottom). The derived  relation (solid black line) has an offset in SFR at a given ${\rm M_{\star}}$ with respect the intrinsic relation (dotted black line) at all redshifts considered (Fig. \ref{fig:SFRSED} and Table \ref{tab_stepsfrf4}, offset$_{z \simeq 4}$ $\sim -0.2$ to $ -0.5 $ dex, offset$_{z \simeq 2}$ $\simeq -0.15 $ to $0$ and offset$_{z \simeq 1}$ $\simeq 0.2$ to $0.5 $ dex) and appears to be flatter at $z \simeq 4$ but steeper at $z \simeq 1$ than the intrinsic slope.  In Appendix \ref{Appendix} we demonstrate that the above is the result of underpredicted SFRs at $z \sim 4$ and underpredicted stellar masses and overpredicted SFRs at $z = 1$. The green squares represent the observations of \citet{Kajisawa2010}, \citet{Bauer11} and \citet{Salmon2015}, while the dashed green lines describe the results of \citet{Pearson2018}. \citet{Kajisawa2010} determined the SFRs of GOODS-North galaxies using dust corrections inferred from SED fitting to the UBV$_{iz}$JHK, 3.6 ${\rm \mu m}$, 4.5 ${\rm \mu m}$, and 5.8 ${\rm \mu m}$ bands alongside with 2800 $ {\rm \AA}$ luminosities and the \citet{kennicutt1998} relation. \citet{Bauer11} derived the SFRs of the GOODS-NICMOS galaxies using their UV luminosities and dust corrections inferred from SED fitting \citep{Calzetti2000,Bruzualch03}. \citet{Salmon2015} retrieved SFRs from the CANDELS and Spitzer Extended Deep Survey.  The authors used a Bayesian SED fitting procedure taking  advantage  of  mock  catalogs  and  synthetic photometry from semi-analytic models. \citet{Pearson2018} obtained the SFRs and stellar masses of the COSMOS galaxies using the CIGALE SED fitting code and assumed  delayed  exponentially declining star formation histories, the \citet{Bruzualch03} stellar population synthesis model and the \citet{Charlot2000} dust  attenuation. The above authors used SED fitting methods to derive properties of galaxies and despite small differences in their assumptions (for more details present see Table \ref{tab_stepsfrf3}) produce similar results. The observational ${\rm SFR}$-${\rm M_{\star}}$ and the EAGLE+SKIRT SFR${\rm_{SED-FAST}}$ $\,$ - $\,$ M${\rm_{\star, SED-FAST}}$ are in good agreement at $z \simeq 1-2 $ but not at redshift $z \simeq 4$ where the SFR${\rm_{SED-FAST}}$-M${\rm_{\star, SED-FAST}}$ relation implies lower values of SFR at fixed stellar mass than observed by $\simeq 0.2 $ to $ 0.5$ dex. Nevertheless, we see already that the assumed methodology to obtain intrinsic properties can have a considerable effect to the derived ${\text{SFR}}$-${\rm M_{\star}}$ relation.

\begin{figure*}
\hspace{-0.20 cm}
\includegraphics[scale=0.35]{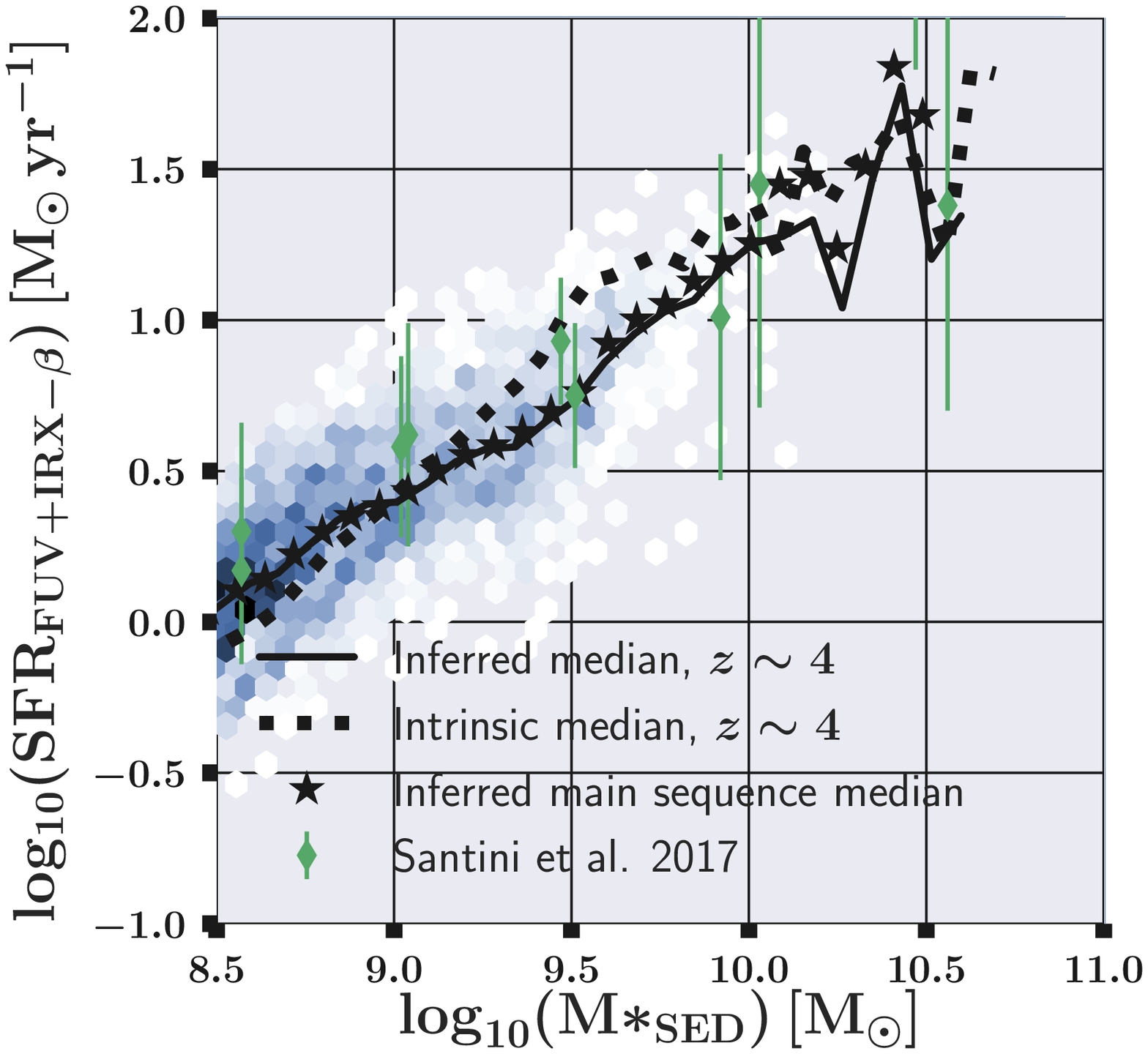}
\hspace{0.20 cm}
\includegraphics[scale=0.35]{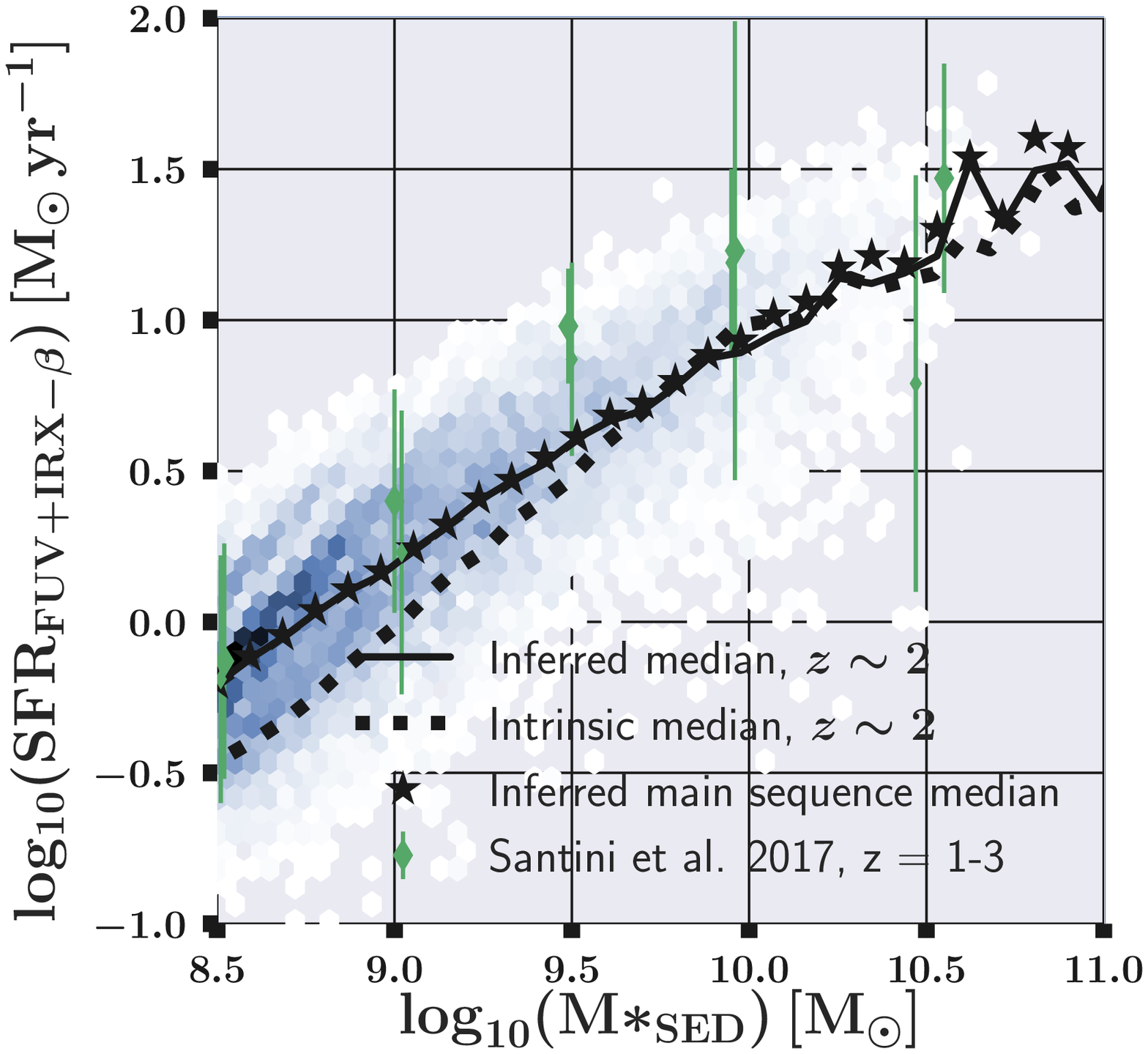}
\hspace{0.20 cm}
\includegraphics[scale=0.35]{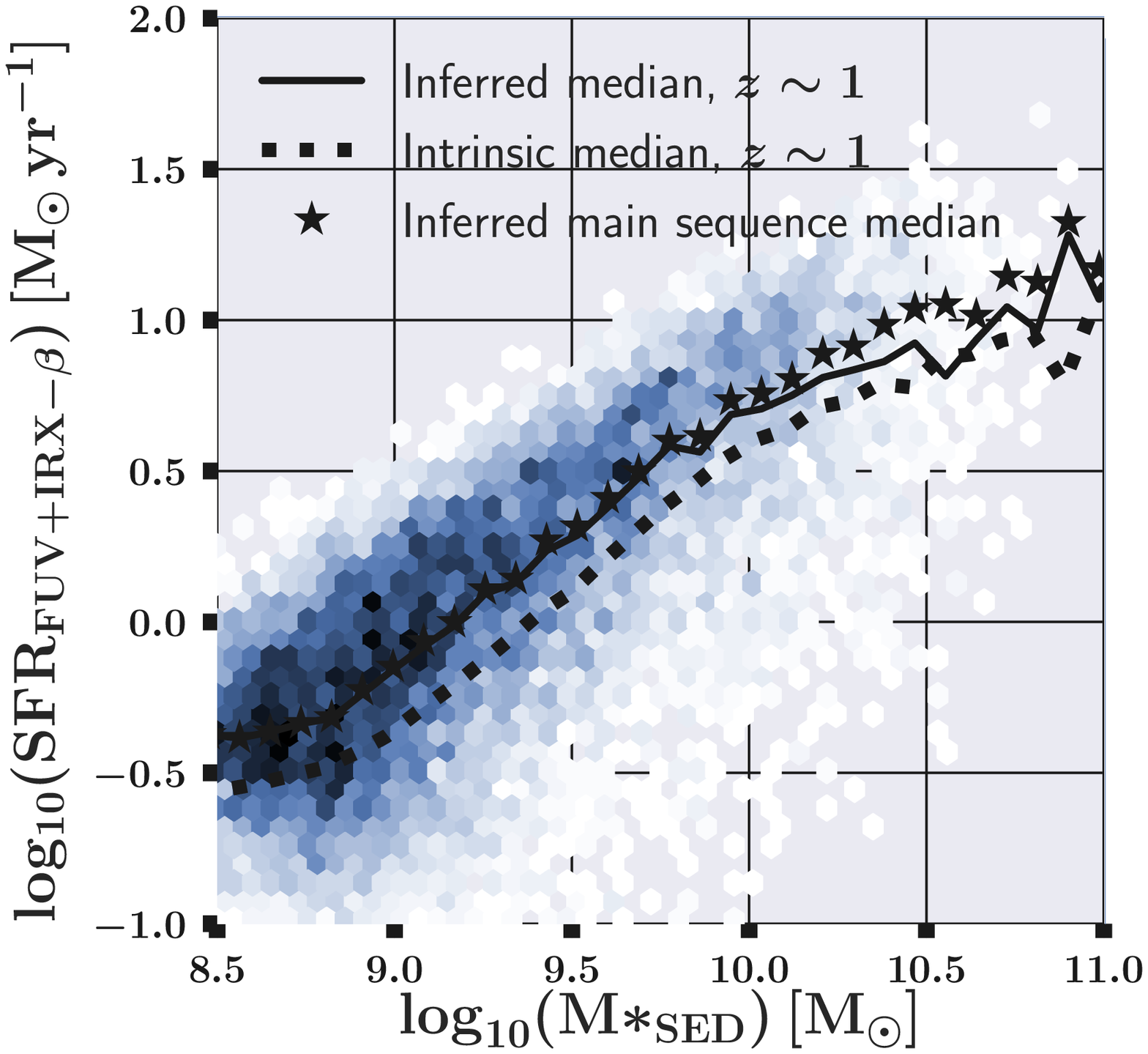} \\
\vspace{0.8 cm}
\includegraphics[scale=0.35]{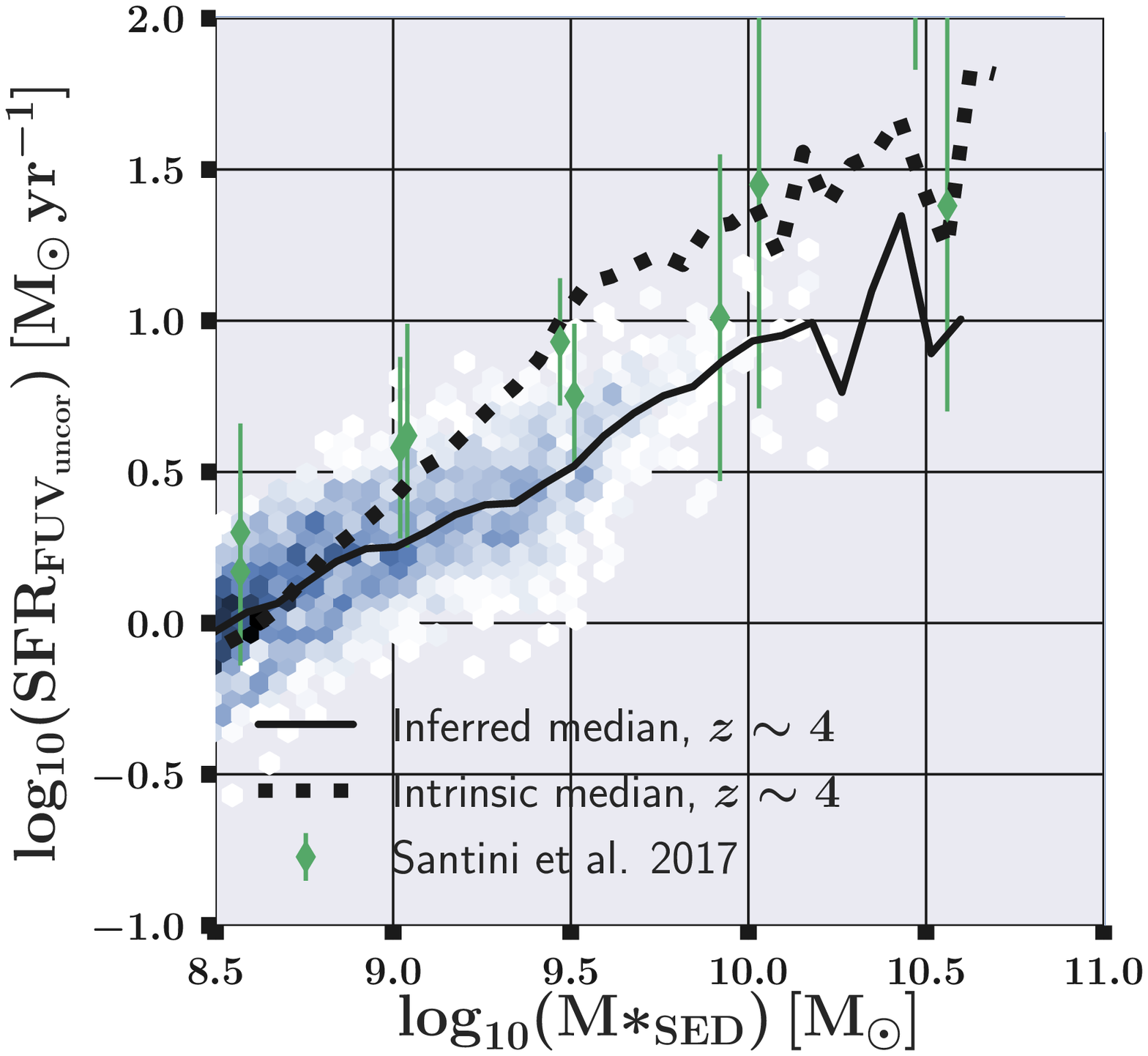}
\hspace{0.20 cm}
\includegraphics[scale=0.35]{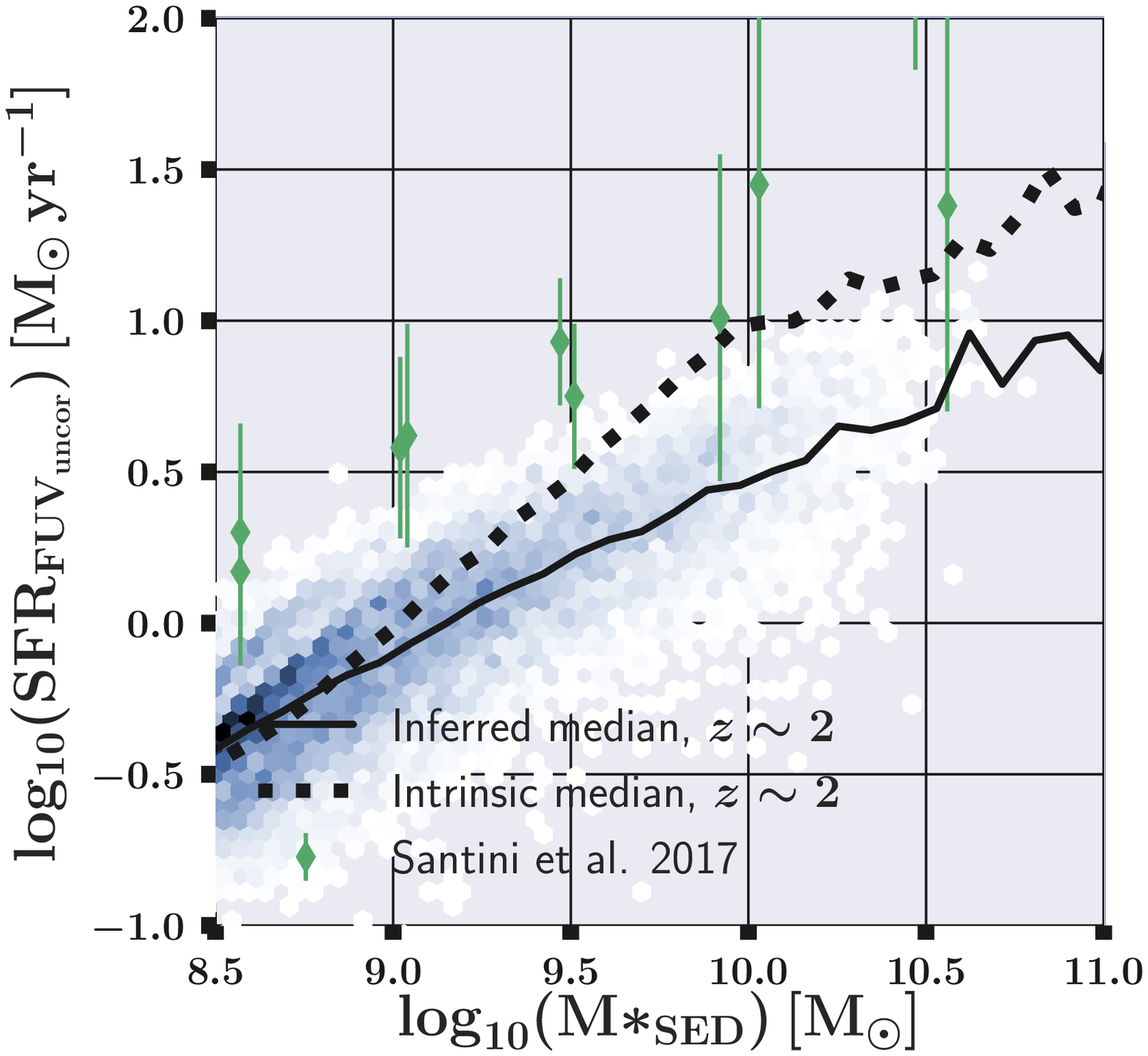}
\hspace{0.20 cm}
\includegraphics[scale=0.35]{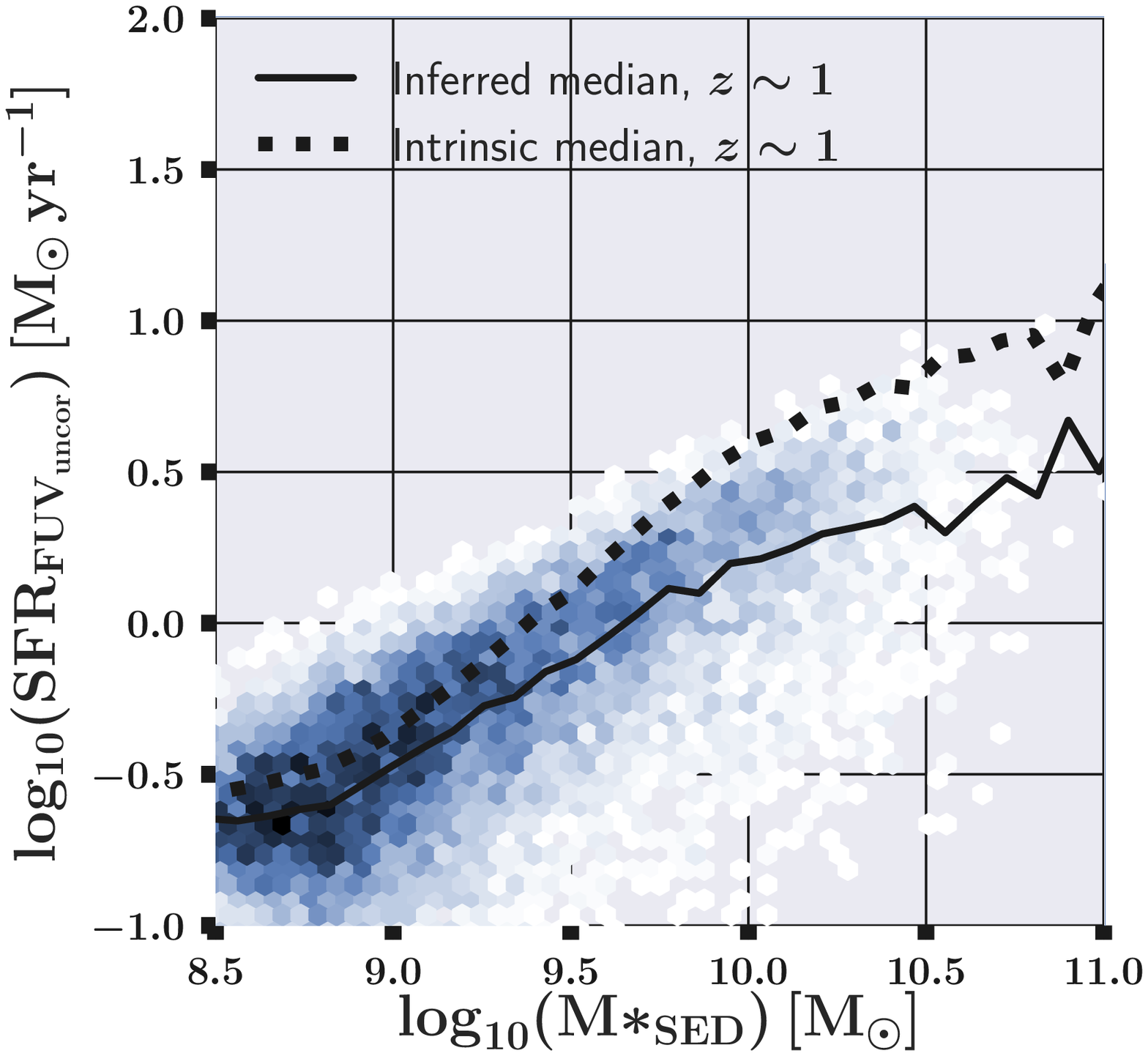}
\caption{Top panels: Same as Fig. \ref{fig:SFRIR}, but for SFRs derived from the FUV luminosity \citep{Kennicutt2012} and the IRX-${\rm \beta}$ relation \citep{meurer1999,bouwens2012,Katsianis2016} while the stellar masses are calculated through the SED fitting technique (black solid line). When applied to the EAGLE+SKIRT data, this method yields a relation which is slightly flatter than the intrinsic (black dotted line). Bottom panels: Same as top but instead dust corrections are not applied.}
\label{fig:SFRUV}
\end{figure*}

\begin{figure}
\includegraphics[scale=0.45]{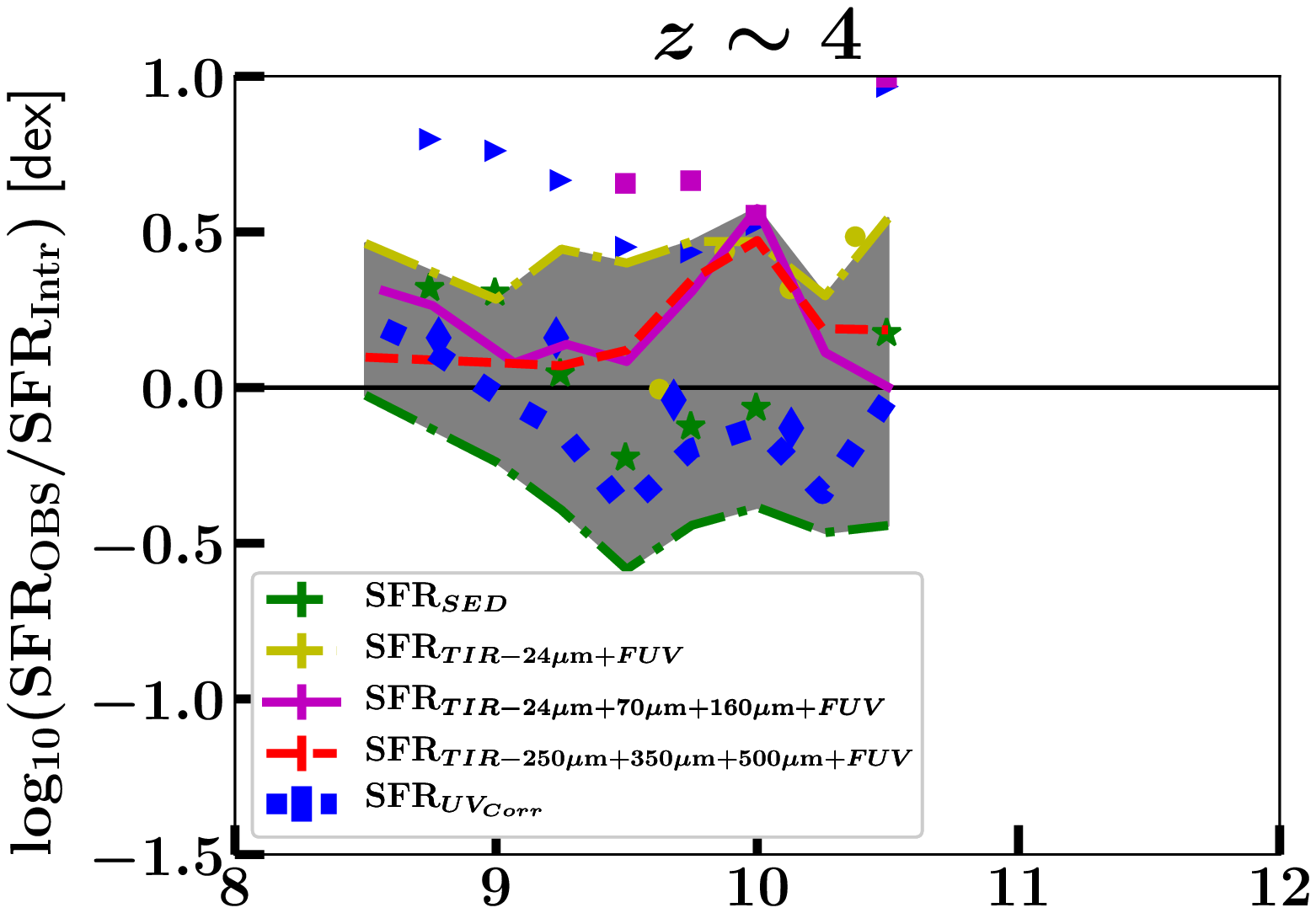}
\includegraphics[scale=0.45]{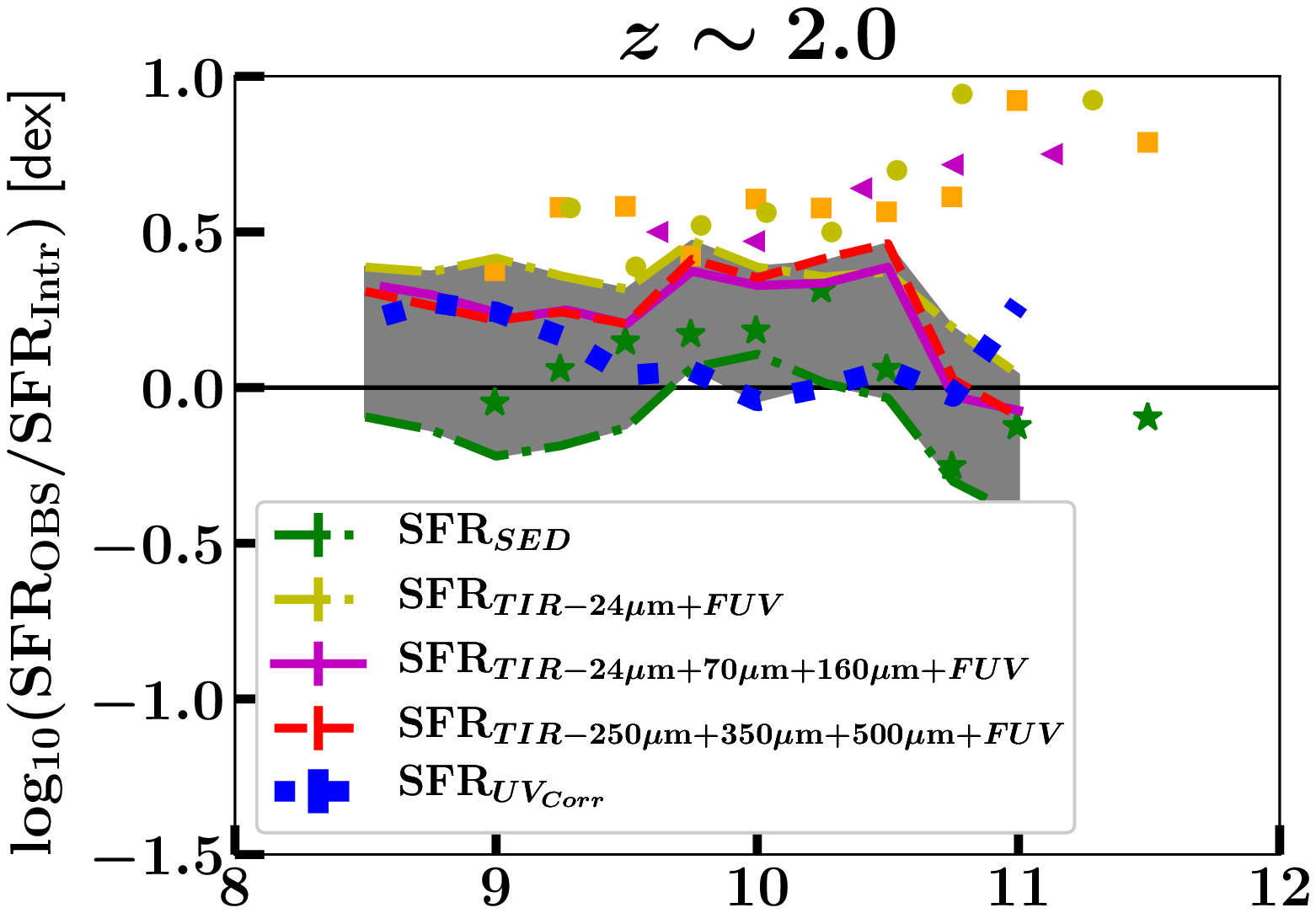}
\includegraphics[scale=0.45]{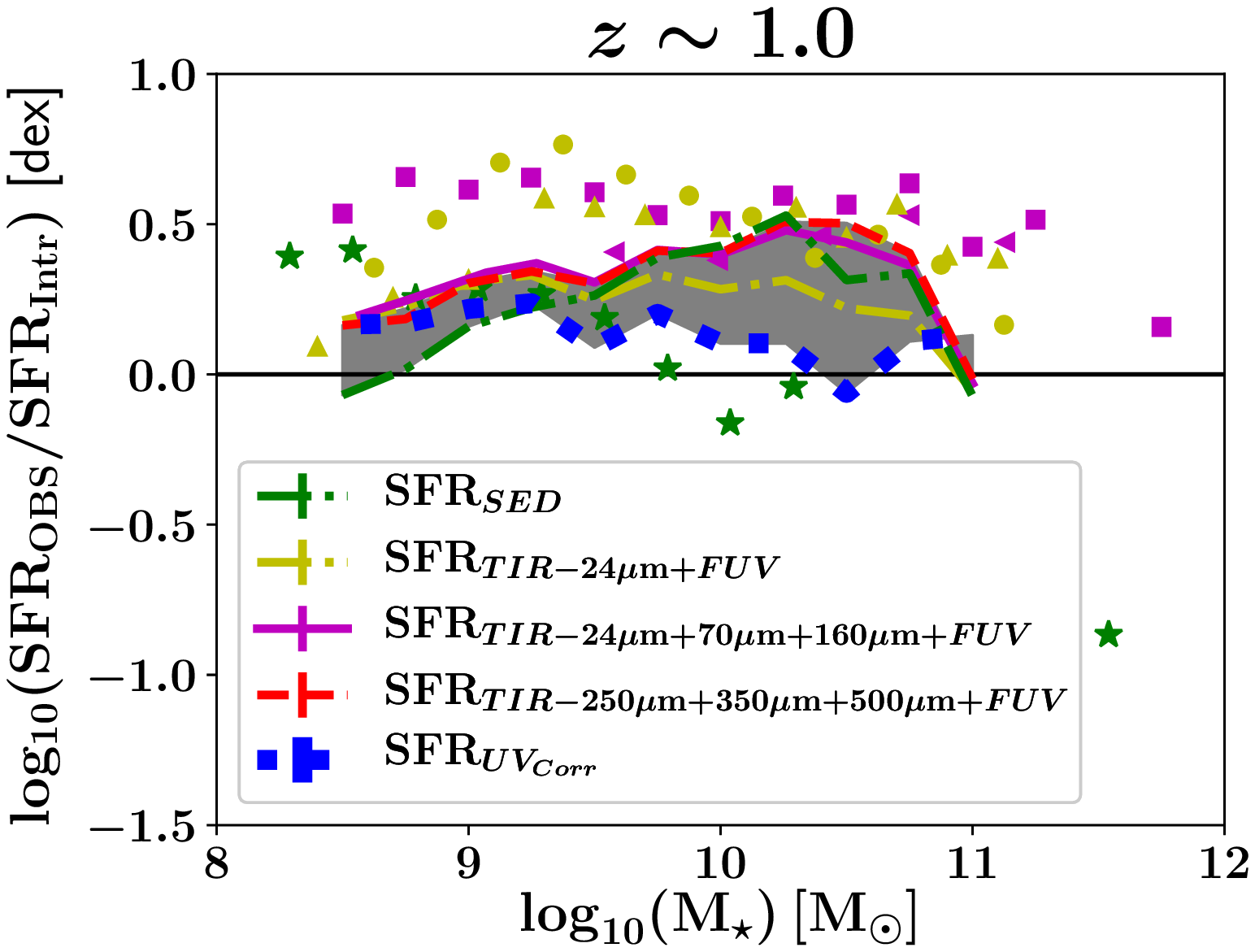}
\vspace{-0.5 cm}
\caption{The offset in dex between the various methods used to derive the ${\rm SFR-M_{\star}}$ relation from the mock EAGLE+SKIRT data with respect the intrinsic EAGLE relation (solid 0 dex line) at $z \simeq 4$ (top), $z \simeq 2$ (medium) and $z \simeq 1$ (bottom). The dark green dashed line represents the offset of the ${\rm SFR-M_{\star}}$ calculated using the FAST SED fitting code.  The orange dash-dotted line represents the SFRs that are inferred from FUV and ${\rm IR_{24 \mu m}}$ luminosities \citep{Wuyts2008}. The magenta solid line represents the results when SFRs are calculated using the 24, 70 and 160 ${\rm \mu m}$ luminosities and the relation given by the  \citet{DaleHelou2002} templates. The red dashed line represents the results when SFRs are calculated using the 250, 350 and 500 ${\rm \mu m}$ luminosities and the \citet{Dale2014} templates. The blue dotted line describes the SFRs derived from UV luminosities  dust-corrected using the IRX-${\rm \beta}$ relation \citep{meurer1999}. The grey area describes the offset in dex between the range of methodologies used in this work which spans areas of $\sim $ 0.5-1 dex at $z = 4$, $\sim$ 0.5 dex at $z = 2$ and $\sim$ 0.1 to 0.5 dex at $z = 1$. We see that the level of discrepancy between different methodologies produced by the EAGLE+SKIRT data resembles that of those observed relations reported in the literature.}
\label{fig:SFRgrey}
\end{figure}

\item  The solid black lines in the left panels of Fig. \ref{fig:SFRIR} represent the SFR${\rm_{24 \mu m - Wuyts \, et \, al. \, 2008}}$ $\,$ - $\,$ M${\rm_{\star, SED-FAST}}$ relation at $z \simeq 4$ (top), $z \simeq 2$ (middle) and $z \simeq 1$ (bottom). The inferred relation (solid black line) is offset to higher SFRs than the intrinsic relation (dotted black line) at all redshifts considered (Fig. \ref{fig:SFRIR} and Table \ref{tab_stepsfrf4}, offset$_{z \simeq 4}$ $\simeq 0.30-0.44 $ dex, offset$_{z \simeq 2}$ $\simeq 0.3 $ and offset$_{z \simeq 1}$ $\simeq 0.25 $ dex). In Appendix \ref{Appendix} we demonstrate that this is the result of underpredicted stellar masses and overpredicted SFRs. The orange squares in the right panels of Fig. \ref{fig:SFRIR} represent the results of \citet{Whitaker2014} who adopted a luminosity-independent conversion from the observed ${\rm IR_{24 \mu m}}$ flux density to the total IR luminosity following \citet{Wuyts2008} and the 2800 $ {\rm \AA}$ emission of 39,106 star forming galaxies selected from the 3D-HST photometric catalogs. The orange circles show the results of \citet{Tomczak2016} who used data from the ZFOURGE survey combined with IR imaging from the Spitzer and Herschel observatories. The authors inferred stellar masses by fitting stellar population synthesis templates \citep{Bruzualch03} to the 0.3-8 ${\rm \mu m}$ photometry using the SED-fitting code FAST \citep{Kriek2009} assuming a \citet{chabrier03} IMF, solar metallicity and exponentially declining star formation histories alongside with a \citet{Calzetti2000} extinction law. SFRs were derived by combining UV and TIR luminosities, where TIR were inferred using the \citet{Wuyts2008} templates. The observational ${\rm SFR}$-${\rm M_{\star}}$ relations are in good agreement with the predicted SFR${\rm_{24 \mu m - Wuyts \, et \, al. \, 2008}}$ $\,$ - $\,$ M${\rm_{\star, SED-FAST}}$ from the EAGLE+SKIRT data. The agreement improves further if a main sequence is specified  (black stars) defined by excluding passive objects imposing a redshift specific star formation rate cut \citep{Schaye2015,Furlong2014,Matthee2018,Katsianis2019}: $sSFR <  10^{-9.1}$ at $z \sim 4$, $sSFR <  10^{-9.6}$ at $z \sim 2$ and $sSFR <  10^{-10.1}$ at $z \sim 1$.  
  
\item  The black solid lines in the middle panels of Fig. \ref{fig:SFRIR} represent the SFR${\rm _{24, 70, 160 \mu m - r \, Dale \& Helou \, 2002}}$-M${\rm_{\star, SED-FAST}}$ relation at $z \simeq 4$ (top), $z \simeq 2$ (middle) and $z \simeq 1$ (bottom) retrieved from the EAGLE+SKIRT data. The dotted black line represents the intrinsic/true relation from the same sample. The inferred relation (solid black line), implies larger SFRs at fixed stellar mass than the intrinsic relation (dotted black line) at all redshifts considered, for masses in the $\log_{10}(M_{\star}/M_{\odot}) \simeq 8.5-10.0$ range (Fig. \ref{fig:SFRIR} and Table \ref{tab_stepsfrf4}, offset$_{z \simeq 4}$ $\simeq 0.1-0.5 $ dex, offset$_{z \simeq 2}$ $\simeq 0.2-0.4 $ and offset $_{z \simeq 1}$ $\simeq 0.2-0.4 $ dex). In the Appendix \ref{Appendix} we demonstrate that this is the result of overpredicted SFRs (by up to 0.3 dex at $z \simeq 2$) and underpredicted stellar masses (by up to -0.20 dex at $z \simeq 2$)\footnote{We note that an overprediction/underprediction of the retrieved SFRs shifts the relation to higher/lower normalizations, while an overprediction/underpredction of stellar masses shifts the ${\rm SFR}$-${\rm M_{\star}}$ relation to lower/higher SFRs at a fixed stellar mass.}. We also plot the observations of \citet[][red dashed lines]{Santini2009}, \citet[][red dotted lines]{Heinis2014} and \citet[][dashed green line]{Steinhartdt2014}. \citet{Santini2009} inferred the TIR of the GOODS-MUSIC galaxies using their 24${\rm \mu m}$ luminosities and the \citet{DaleHelou2002} templates and combined the above TIR luminosities with UV emission ($2700 {\rm \AA}$) in order to derive the galaxy SFRs. \citet{Steinhartdt2014} used the far-infrared  Herschel wavelengths and employed the \citet{Casey2012} models which are very similar to the \citet{DaleHelou2002} templates. The SFR${\rm_{24, 70, 160 \mu m - r \, Dale \& Helou \, 2002}}$ $\,$ - $\,$ M${\rm_{\star, SED-FAST}}$ relation derived from EAGLE+SKIRT simulations is in agreement with observations. 

\item The black solid lines in the right panels of Fig. \ref{fig:SFRIR} represent the SFR${\rm _{250, 350, 500 \mu m - c \, Dale \& Helou \, 2014}}$-M${\rm_{\star, SED-FAST}}$ relation at $z \simeq 4$ (top), $z \simeq 2$ (middle) and $z \simeq 1$ (bottom) retrieved from the EAGLE+SKIRT data. Similarly with the middle panel, in which the SFR${\rm _{24, 70, 160 \mu m - r \, Dale \& Helou \, 2002}}$-M${\rm_{\star, SED-FAST}}$ is described,  the inferred relation (solid black line) implies larger SFRs at fixed stellar mass than the intrinsic relation (dotted black line) at all redshifts considered, for masses in the $\log_{10}(M_{\star}/M_{\odot}) \simeq 8.5-10.5$ range (Fig. \ref{fig:SFRIR} and Table \ref{tab_stepsfrf4}).

\item The top black solid lines in Fig. \ref{fig:SFRUV} represent the SFR${\rm_{UV+IRX-\beta}}$ $\,$ - $\,$ M${\rm_{\star, SED-FAST}}$ relation at $z \simeq 4$ (left), $z \simeq 2$ (middle) and $z \simeq 1$ (right)\footnote{In order to demonstrate the effect of dust-corrections we present the relation if UV light is dust uncorrected at the bottom panel.}. The derived relation (black solid line) has an offset with respect to the intrinsic relation (black dotted line) of offset$_{z \simeq 4}$ $\simeq 0.11$ to $-0.13 $ dex, offset$_{z \simeq 2}$ $\simeq 0.23$  to $-0.02$ and offset$_{z \simeq 1}$ $\simeq 0.22 $ to $-0.08$ dex (Fig. \ref{fig:SFRUV} and Table \ref{tab_stepsfrf4}). At $z \sim 4$ for masses in the  $\log_{10}(M_{\star}/M_{\odot}) \simeq 8.5-9$ range the SFRs are typically overestimated. However, the SFRs are underestimated for  $\log_{10}(M_{\star}/M_{\odot}) > 9.5$. This makes the inferred SFR${\rm_{UV+IRX-\beta}}$ $\,$ - $\,$ M${\rm_{\star, SED-FAST}}$ relation flatter. \citet{Santini2017} inferred the ${\rm SFR}$-${\rm M_{\star}}$ relation for the HST Frontier fields galaxies, based on rest-frame UV observations, the \citet{Kennicutt2012} relation and the \citet{meurer1999} dust correction law. We see that both the derived SFR${\rm_{UV+IRX-\beta}}$ $\,$ - $\,$ M${\rm_{\star, SED-FAST}}$ (black solid line) and SFR${\rm_{Intr}}$ $\,$ - $\,$ M${\rm_{\star, Intr}}$ (black dotted line) relations are consistent with the observations. A common finding for all redhifts of interest is that the derived relation is flatter than the intrinsic.

\end{itemize}

In Fig. \ref{fig:SFRgrey} we present the offset in dex  with respect the intrinsic/true EAGLE+SKIRT relation for all methodologies used to derive  the ${\rm SFR-M_{\star}}$ relation from the EAGLE+SKIRT data  at $z \simeq 4$ (top), $z \simeq 2$ (middle) and $z \simeq 1$ (bottom). The dark green dot-dashed line represents the offset of the ${\rm SFR-M_{\star}}$ calculated using the FAST SED fitting code.  The orange dash-dotted line represents the SFRs that are inferred from FUV and ${\rm IR_{24 \mu m}}$ luminosities \citep{Wuyts2008}. The magenta solid line represents the SFR${\rm_{24, 70, 160 \mu m - r \, Dale \& Helou \, 2002}}$ $\,$ - $\,$ M${\rm_{\star, SED-FAST}}$ vs SFR${\rm_{Intr}}$ $\,$ - $\,$ M${\rm_{Intr}}$ relation. The red rashed line represents the SFR${\rm_{250, 350, 500 \mu m - c \, Dale \& Helou \, 2014}}$ $\,$ - $\,$ M${\rm_{\star, SED-FAST}}$ vs SFR${\rm_{Intr}}$ $\,$ - $\,$ M${\rm_{Intr}}$ relation. The blue dotted line represents the SFR${\rm _{UV+IRX-\beta}}$ $\,$ - $\,$ M${\rm_{\star, SED-FAST}}$ relation. The grey area encompaces the offset between the range of different methodologies used in this work. The results span areas of $\sim $ 0.5 to 1.0 dex at $z \sim 4 $, 0.5 dex at $z \sim 2$ and 0.1 to 0.5 dex at $z \sim 1$. Alongside we present the observed relations shown in Fig. \ref{fig:SFRIR1} in order to demonstrate that a similar level of tension exists between them. Thus, considering the comparisons present at figures \ref{fig:SFRSED}, \ref{fig:SFRIR},  \ref{fig:SFRUV} and \ref{fig:SFRgrey}, we suggest that the discrepancies between observational studies have largely their roots in the diversity of methodologies used in the literature to derive SFRs \citep{Katsianis2015}. We note that the tension  represented by the grey area reported above, reproduced by the EAGLE+SKIRT data, has its roots solely in differences in SFR determinations since stellar masses are in all cases computed with the same technique. A further future analysis which explores selection effects to the ${\rm SFR-M_{\star}}$ relations employing mock observations can probably be used to supplementary address the tension between observations in the literature.

\section{Discussion and conclusions}
\label{Disc}

\begin{table*}
\centering
\resizebox{0.95\textwidth}{!}{%
  \begin{tabular}{ccccccccc}
    \hline \\
    & {\large Methodology}  &
           {\Large $10^{8.5}$} & {\Large $10^{9.0}$} & {\Large $10^{9.5}$} & {\Large $10^{10}$} & {\Large $10^{10.5}$} & {\Large $10^{11.0}$}\\
     &  &  & & Offset  & (dex)        \\
         \hline  \hline
           & SFR${\rm_{SED-FAST}}$ $\,$ - $\,$ M${\rm_{\star, SED-FAST}}$, z = 4   & -0.10 & -0.30 & -0.58 & -0.38 & -0.35 & - \\
    & SFR${\rm_{SED-FAST}}$ $\,$ - $\,$ M${\rm_{\star, SED-FAST}}$, z = 2   & -0.10 & -0.12 & -0.06 & -0.01 & -0.01 & -0.04 \\
    & SFR${\rm_{SED-FAST}}$ $\,$ - $\,$ M${\rm_{\star, SED-FAST}}$, z = 1   & -0.01 & 0.16 & 0.26 & 0.43 & 0.56 & 0.07 \\
         \hline
    & SFR${\rm_{24 \mu m - Wuyts \, et \, al. \, 2008}}$ $\,$ - $\,$ M${\rm_{\star, SED-FAST}}$, z = 4   & 0.46 & 0.37 & 0.31 & 0.46 & 0.54 & - \\ 
    & SFR${\rm_{24 \mu m - Wuyts \, et \, al. \, 2008}}$ $\,$ - $\,$ M${\rm_{\star, SED-FAST}}$, z = 2   & 0.38 & 0.32 & 0.32 & 0.39 & 0.37 & 0.04 \\
         & SFR${\rm_{24 \mu m - Wuyts \, et \, al. \, 2008}}$ $\,$ - $\,$ M${\rm_{\star, SED-FAST}}$, z = 1   & 0.18 & 0.31 & 0.34 & 0.28 & 0.22 & -0.02 \\
          \hline
    & SFR${\rm_{24, 70, 160 \mu m - r \, Dale \& Helou \, 2002}}$ $\,$ - $\,$ M${\rm_{\star, SED-FAST}}$, z = 4   & 0.31 & 0.09 & 0.10 & 0.49 & -0.01 & - \\
    & SFR${\rm_{24, 70, 160 \mu m - r \, Dale \& Helou \, 2002}}$ $\,$ - $\,$ M${\rm_{\star, SED-FAST}}$, z = 2  & 0.32 & 0.32 & 0.20 & 0.33 & 0.38 & -0.01 \\
    & SFR${\rm_{24, 70, 160 \mu m - r \, Dale \& Helou \, 2002}}$ $\,$ - $\,$ M${\rm_{\star, SED-FAST}}$, z = 1  & 0.19 & 0.34 & 0.30 & 0.40 & 0.43 & -0.02 \\
          \hline
          & SFR${\rm_{250, 350, 500 \mu m - c \, Dale \& Helou \, 2014}}$ $\,$ - $\,$ M${\rm_{\star, SED-FAST}}$, z = 4   & 0.22 & 0.16 & 0.11 & 0.47 & 0.06 & - \\
    & SFR${\rm_{250, 350, 500 \mu m - c \, Dale \& Helou \, 2014}}$ $\,$ - $\,$ M${\rm_{\star, SED-FAST}}$, z = 2  & 0.22 & 0.31 & 0.20 & 0.35 & 0.46 & -0.07 \\
    & SFR${\rm_{250, 350, 500 \mu m - c \, Dale \& Helou \, 2014}}$ $\,$ - $\,$ M${\rm_{\star, SED-FAST}}$, z = 1  & 0.16 & 0.31 & 0.40 & 0.40 & 0.56 & -0.01 \\
    \hline
    &  SFR${\rm_{UV+IRX-\beta}}$ $\,$ - $\,$ M${\rm_{\star, SED-FAST}}$, z = 4  & 0.11 & -0.02 & -0.13 & -0.04 & -0.02 & - \\
    &  SFR${\rm_{UV+IRX-\beta}}$ $\,$ - $\,$ M${\rm_{\star, SED-FAST}}$, z = 2  & 0.23 & 0.16 & 0.04 & -0.02 & -0.01 & -0.02 \\
    &  SFR${\rm_{UV+IRX-\beta}}$ $\,$ - $\,$ M${\rm_{\star, SED-FAST}}$, z = 1  & 0.17 & 0.22 & 0.19 & 0.10 & -0.02 & -0.08 \\
\hline \hline
  \end{tabular}%
}
\caption{The offset in dex between the derived and intrinsic SFR-M${\rm_{\star}}$ relations at different masses. To infer the intrinsic relation a decrement equal to the offset we report is required.}
\label{tab_stepsfrf4}
\end{table*}

Significant tension has been reported between observed high-redshift star formation rate (SFR) - stellar mass (${\rm M}_{\star}$) relations reported by different authors in terms of normalization, shape and slope (section \ref{thecode}). We examined the SFR$-{\rm M}_{\star}$ relation of $z \simeq 1-4 $ galaxies using the SKIRT simulated spectral energy distributions \citep{Camps2018} from the EAGLE hydrodynamic simulations. We derived SFRs and stellar masses using different observational techniques (e.g. SED fitting, UV+TIR luminosities, IR${\rm_{24}}$ data and UV+IRX-${\rm \beta}$ relation). We compared our results from the simulated data with a range of observed relations and revisited the inconsistency reported between observed and simulated SFR-M* relations in the literature \citep[e.g.][]{Sparre2014,Katsianis2015}. Our main findings are:

\begin{itemize}

\item The tension between the observed and simulated SFR$-{\rm M}_{\star}$ relations at $z \simeq 1-4 $ can be largely alleviated. The discrepancy is decreased considerably when methodological biases, associated with estimating SFR and $M_{\star}$ from observations, are taken into account (Section \ref{thecode3},  Fig. \ref{fig:SFRSED}, \ref{fig:SFRIR} and \ref{fig:SFRUV}).

\item SFRs derived from combinations of Infrared wavelengths (e.g. 24 ${\rm \mu m}$, 24, 70 and 160 ${\rm \mu m}$ or 250, 350, 500 ${\rm \mu m}$) with UV luminosities are significantly overestimated with respect to the intrinsic values by 0.2-0.5 dex (at $z \simeq 1-4$) for the $\log_{10}(M_{\star}/M_{\odot}) \simeq 8.5-10.5$ range. The above results in significantly high normalizations for the SFR${\rm_{24 \mu m - Wuyts \, et \, al. \, 2008}}$ $\,$ - $\,$ M${\rm_{\star, SED-FAST}}$, SFR${\rm_{24, 70, 160 \mu m - r \, Dale \& Helou \, 2002}}$ $\,$ - $\,$ M${\rm_{\star, SED-FAST}}$ and SFR${\rm_{250, 350, 500 \mu m - c \, Dale \& Helou \, 2014}}$ $\,$ - $\,$ M${\rm_{\star, SED-FAST}}$. On the other hand,  SFR${\rm _{UV+IRX-\beta}}$ $\,$ - $\,$ M${\rm_{\star, SED-FAST}}$ relations that rely on SFRs inferred solely from dust corrected UV luminosities are flatter with deviations from the intrinsic values of up to -0.13 dex at $z \simeq 4$. We find that the normalization of SFR${\rm_{SED-FAST}}$ $\,$ - $\,$ M${\rm_{\star, SED-FAST}}$ is significantly underestimated by up to -0.58 dex at $z \simeq 4$ but overestimated by up to 0.3 dex at $z \sim 1$ (Section \ref{thecode3}, Fig. \ref{fig:SFRSED}, \ref{fig:SFRIR},  \ref{fig:SFRUV}). 

\item The tension between different observational studies (up to 0.8 dex at $z \simeq 4$ and up to 0.5 dex at $z \simeq 1$, subsection \ref{thecode1}) is at a great extent driven by the different techniques used by different groups to derive observational SFRs (Section \ref{thecode3}, Fig. \ref{fig:SFRgrey}) with significant redshift dependence on the level of mis-estimation.

\end{itemize}

\section*{Acknowledgments}

This work used the DiRAC Data Centric system at  Durham  University,  operated  by  the  Institute for  Computational  Cosmology  on  behalf  of  the  STFC  DiRAC  HPC  Facility  (www.dirac.ac.uk). A.K has been supported by the {\it Tsung-Dao Lee Institute Fellowship}, {\it Shanghai Jiao Tong University} and {\it CONICYT/FONDECYT fellowship, project number: 3160049}. X.Y. is supported by the  national science foundation of China (grant Nos. 11833005, 11890692, 11621303) and Shanghai Natural Science Foundation, grant No. 15ZR1446700. We also thank the support of the Key Laboratory for Particle Physics, Astrophysics and Cosmology, Ministry of Education. C.L. has received funding from the ARC Centre of Excellence for All Sky Astrophysics in 3 Dimensions (ASTRO 3D), through project number CE170100013. Cosmic Dawn Centre is funded by the Danish National Research Foundation. M.S. acknowledges support by the Ministry of Education, Science, and Technological Development of the Republic of Serbia through the projects Astrophysical Spectroscopy of Extragalactic Objects (176001) and Gravitation and the Large Scale Structure of the Universe (176003). We would like to thank the anonymous referee for their suggestions and comments which improved significanlty our work.


\bibliographystyle{mn2e}	
\bibliography{Katsianis_mnrasRev.bbl}

\appendix

\section{Comparison between intrinsic and inferred SFRs and stellar masses}
\label{Appendix}

\begin{figure*}
  \centering
\vspace{0.45 cm}
\hspace{-0.70 cm}
\includegraphics[scale=0.35]{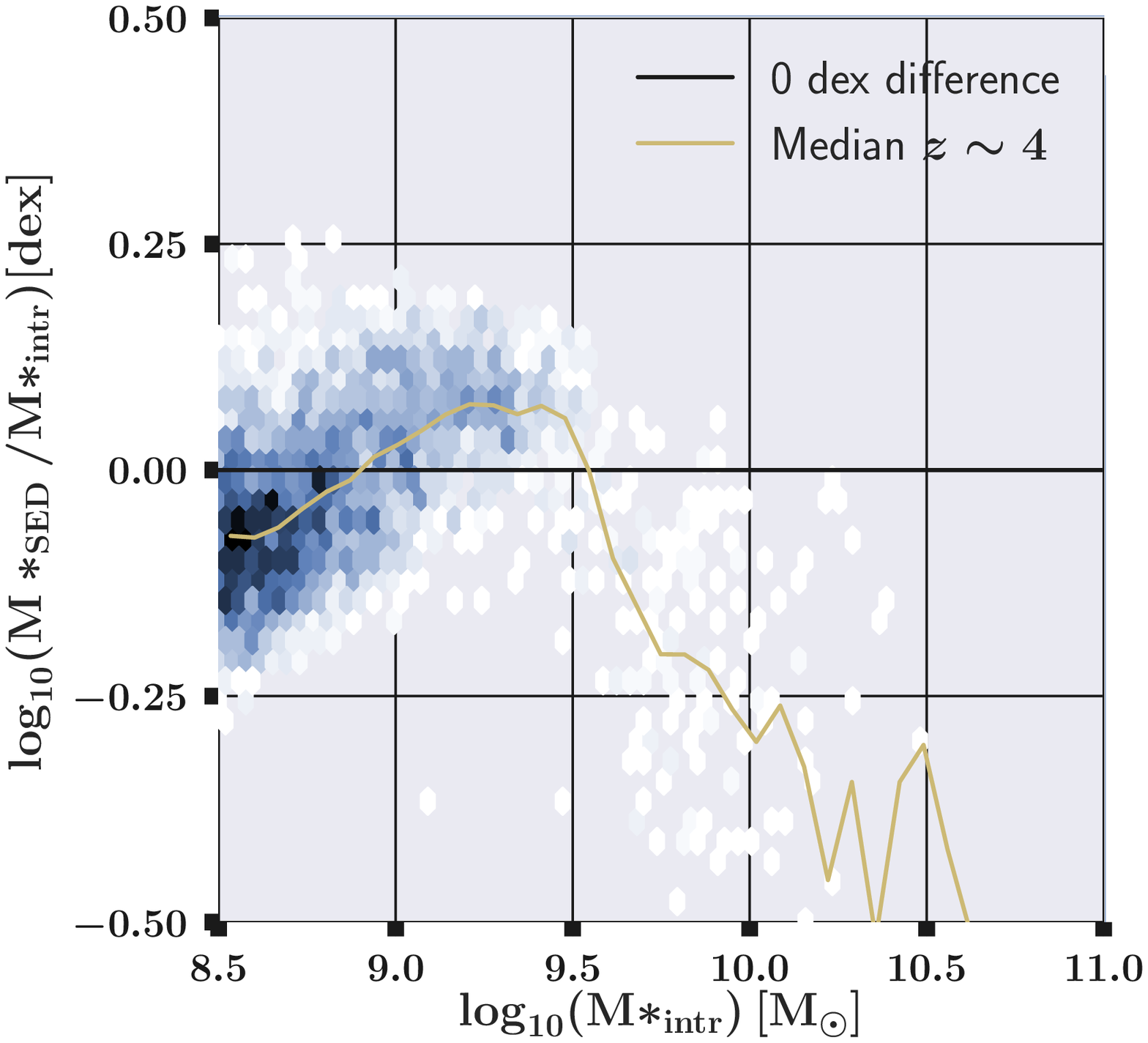}
\hspace{0.30 cm}
\includegraphics[scale=0.35]{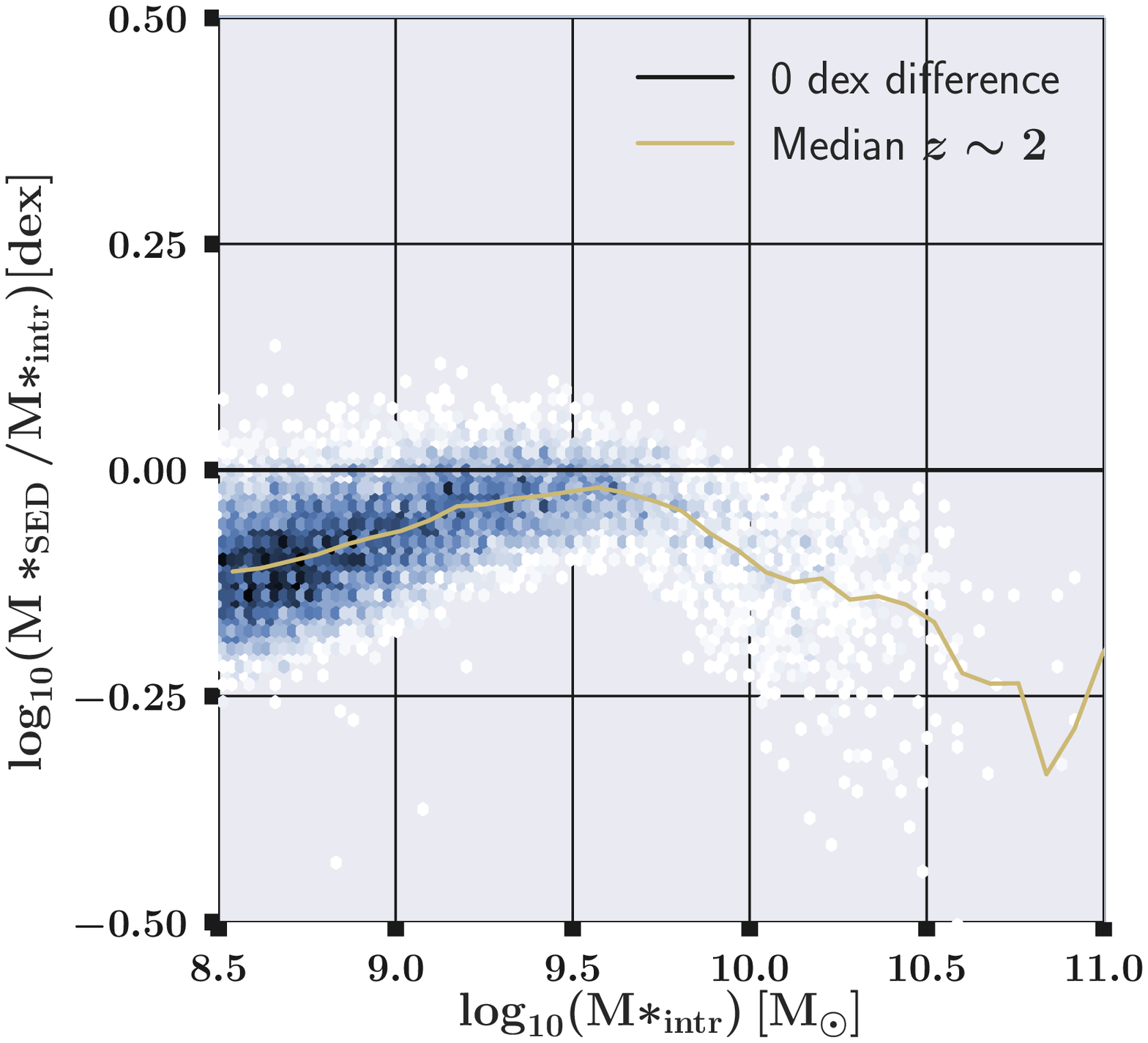}
\hspace{0.30 cm}
\includegraphics[scale=0.35]{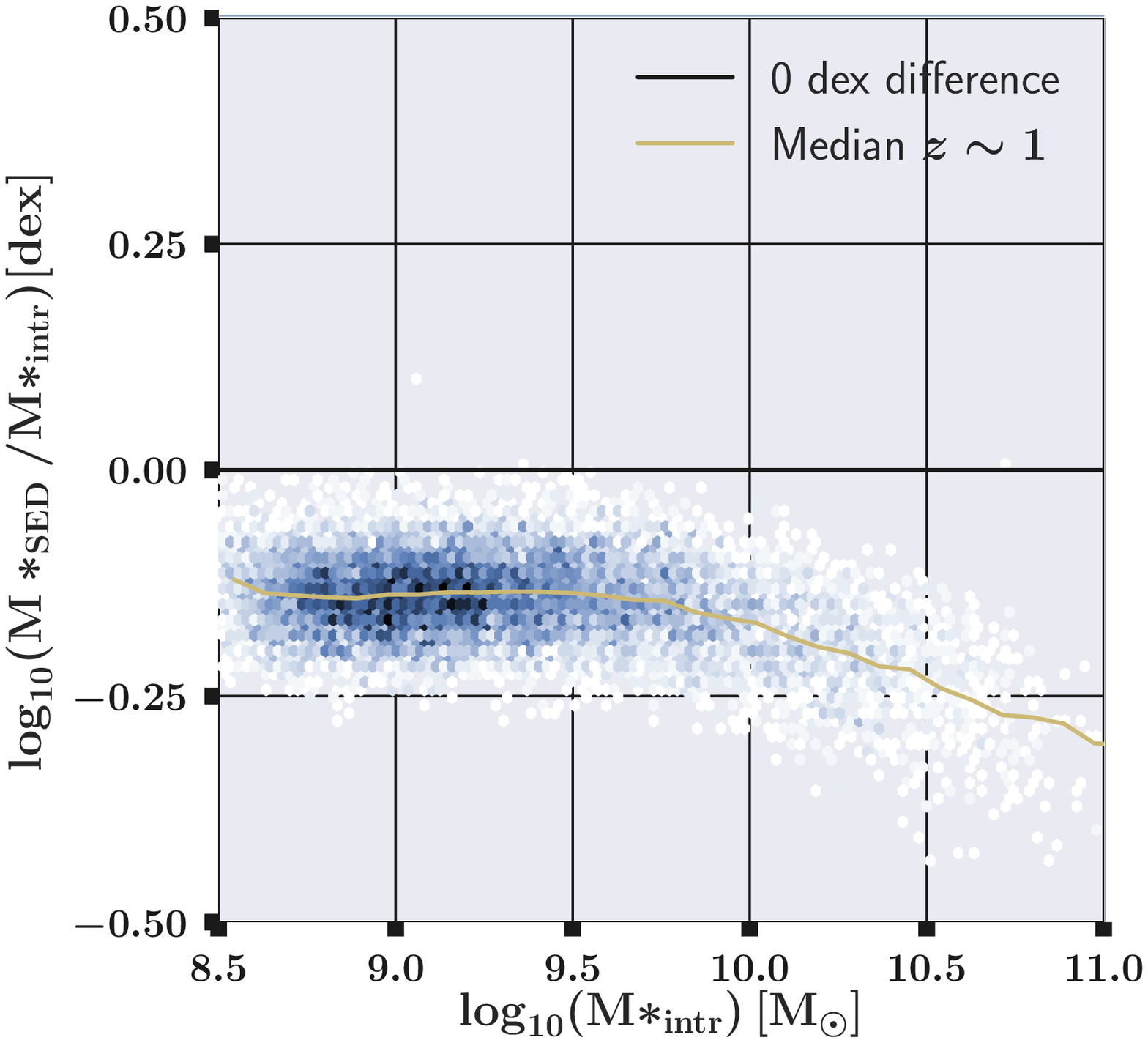} \\
\vspace{0.5 cm}

\includegraphics[scale=0.35]{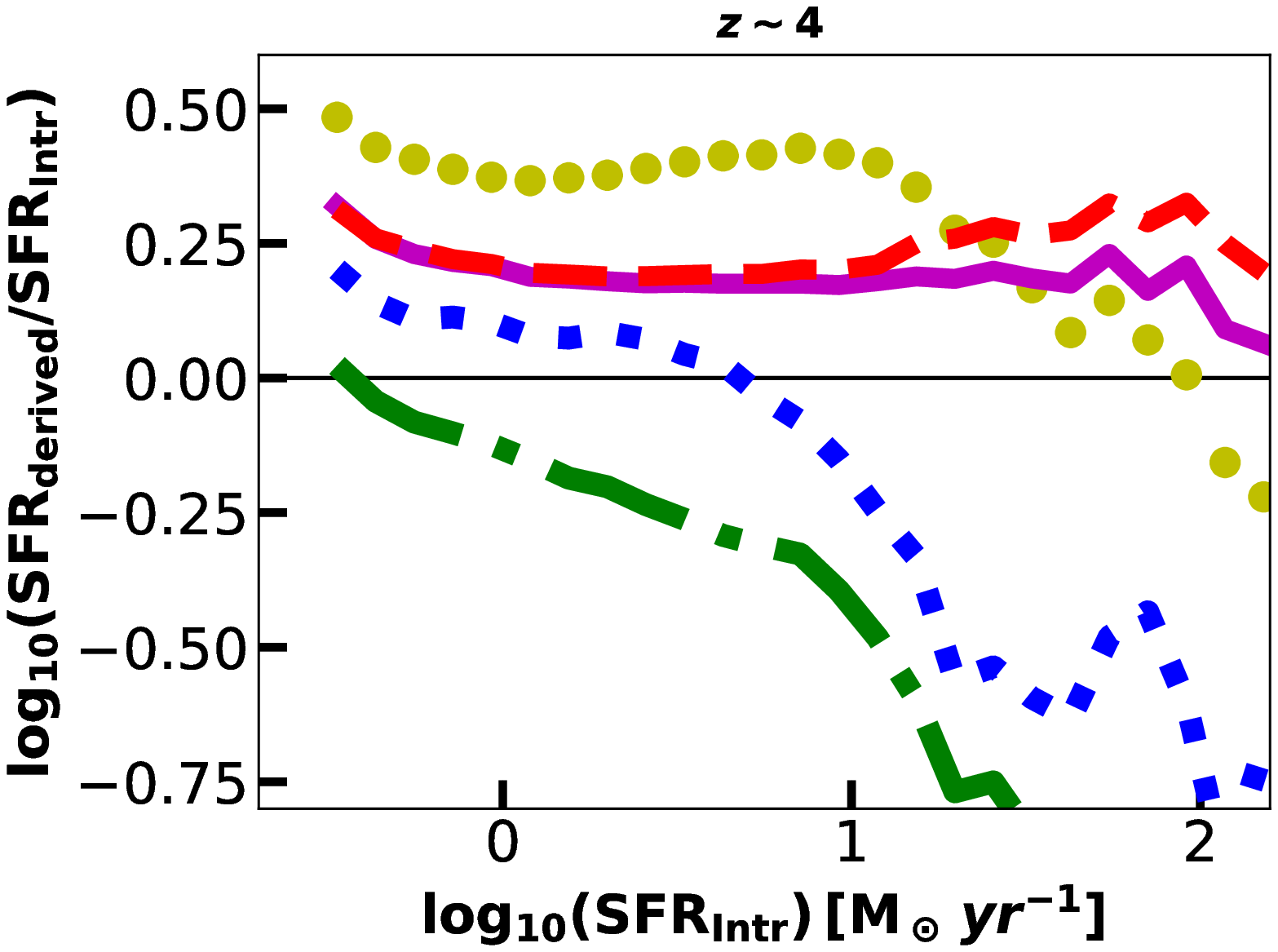}
\includegraphics[scale=0.35]{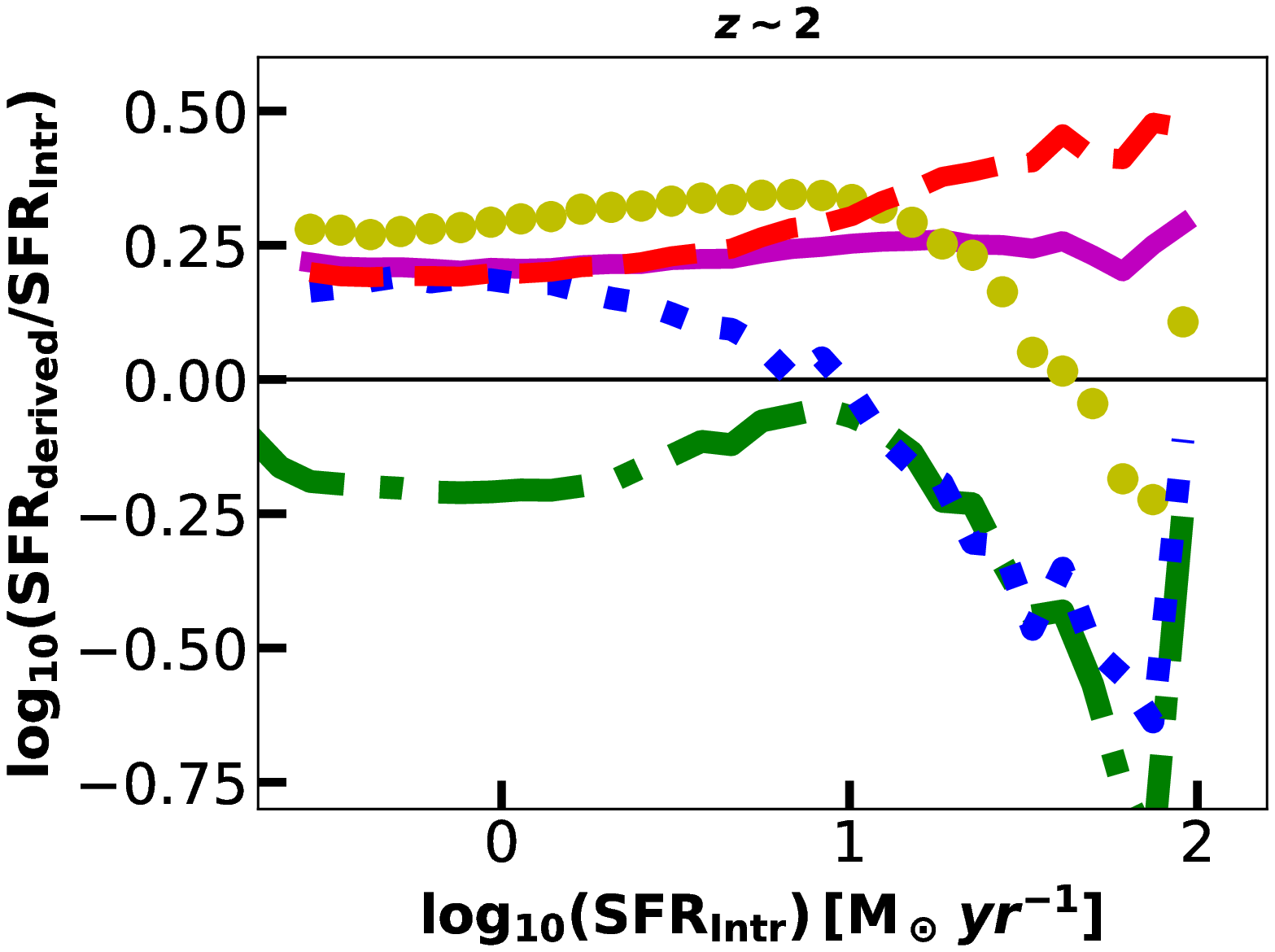}
\includegraphics[scale=0.35]{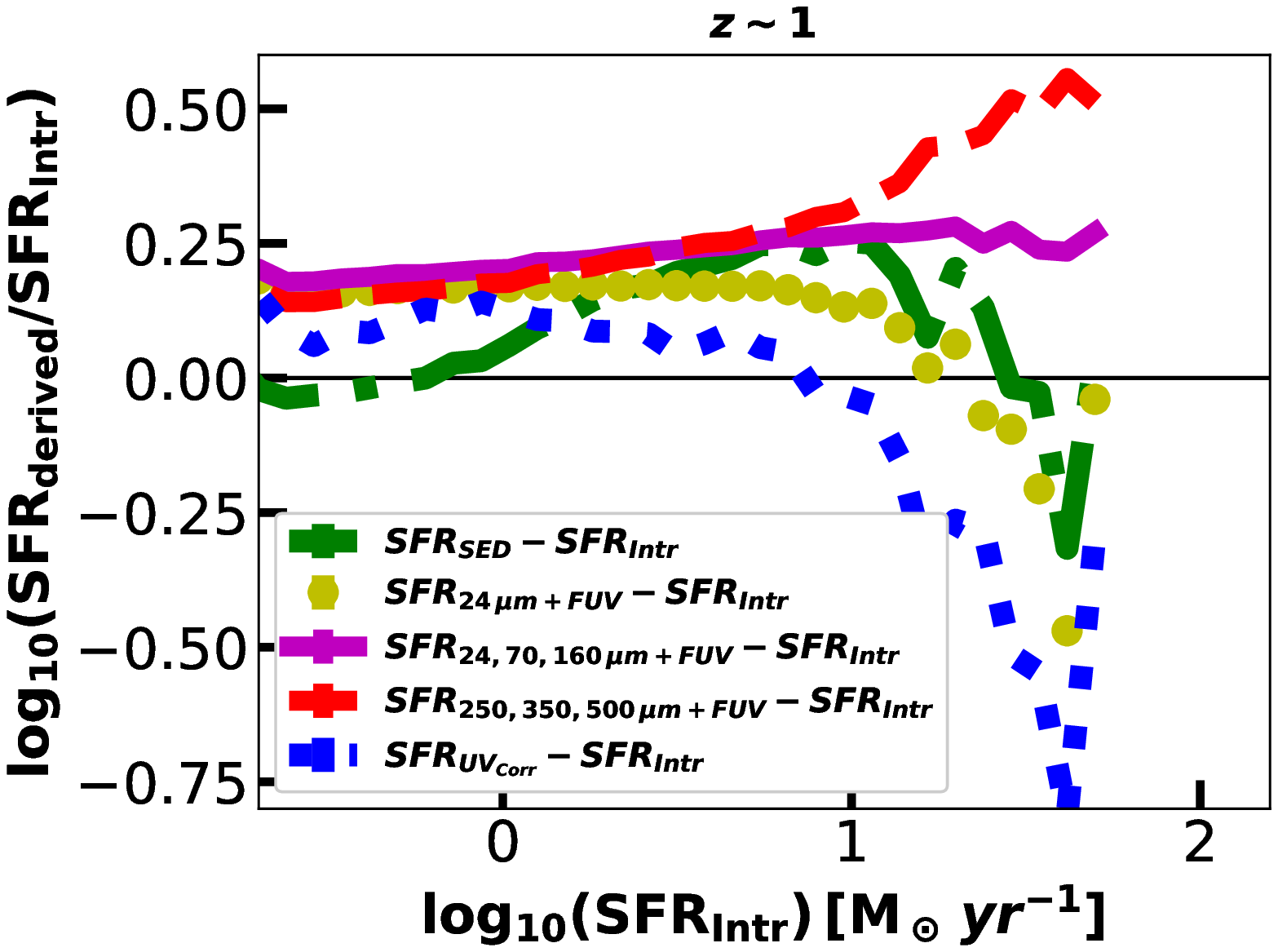}
\caption{Top panels: Offset in dex between M${\rm_{\star, SED-FAST}}$ and ${\rm M_{\star, Intr}}$. Bottom panels: Offset between the SFR${\rm_{24, 70, 160 \mu m - r \, Dale \& Helou \, 2002}}$, SFR${\rm_{SED-FAST}}$, SFR${\rm_{24 \mu m - Wuyts \, et \, al. \, 2008}}$, SFR${\rm_{UV+IRX-\beta}}$ and SFR${\rm_{Intr}}$}
\label{AppendixFig}
\end{figure*}

\begin{table*}
\centering
\resizebox{0.75\textwidth}{!}{%
  \begin{tabular}{ccccccccc}
    \hline \\
    & {\large Methodology}  &
           {\Large $8.5$} & {\Large $9.0$} & {\Large $9.5$} & {\Large $10.0$} & {\Large $10.5$} & {\Large $11.0$}\\
     &  &  & & Offset  & (dex)        \\
    \hline \hline
    &  M${\rm_{\star, SED-FAST}}$, z = 4  & -0.07 & 0.03 & 0.0 & -0.26 & -0.30 & - \\
    &  M${\rm_{\star, SED-FAST}}$, z = 2  & -0.11  & -0.06 & -0.02 & -0.09 & -0.17 & -0.20 \\
    &  M${\rm_{\star, SED-FAST}}$, z = 1  & -0.12 & -0.14 & -0.14 & -0.16  & -0.24 & -0.30 \\
\hline \hline
  \end{tabular}%
}
\caption{The offset in dex between the inferred and intrinsic stellar masses at a fixed intrinsic stellar mass.}
\label{tab_stepsfrf5}
\end{table*}

In this appendix we compare the  SFR${\rm_{SED-FAST}}$,  SFR${\rm_{24 \mu m - Wuyts \, et \, al. \, 2008}}$,  SFR${\rm_{24, 70, 160 \mu m - r \, Dale \& Helou \, 2002}}$, SFR${\rm_{250, 350, 500 \mu m - c \, Dale \& Helou \, 2014}}$,  SFR${\rm_{UV+IRX-\beta}}$ and M${\rm_{SED-FAST}}$ calculated from the mock EAGLE+SKIRT galaxies as described in section \ref{thecode2} to the intrinsic SFR${\rm_{Intr}}$ and M${\rm_{\star, Intr}}$ provided in the EAGLE database. In the top panels of Fig. \ref{AppendixFig} and table \ref{tab_stepsfrf5} we present the offset in dex between the M${\rm_{\star, SED-FAST}}$ retrieved from the FAST SED fitting code \citep{Kriek2009} and the intrinsic stellar masses M${\rm_{\star, Intr}}$. We show that at $z \simeq 4$ (top left panel of Fig. \ref{AppendixFig}) the offset between the  M${\rm_{\star, SED-FAST}}$ and M${\rm_{\star, Intr}}$ is $-0.1$ to 0.1  dex in the $\log_{10}(M_{\star}/M_{\odot}) = 8.5 - 10$ range. The  M${\rm_{\star, SED-FAST}}$/M${\rm_{\star, intr}}$ ratio reaches -0.3 at $\log_{10}(M_{\star}/M_{\odot}) = 10.5$ at $z \simeq 4$ (top left panel of Fig. \ref{AppendixFig}). In the middle panel of Fig. \ref{AppendixFig} we demonstrate that the offset is $-0.1$ to $-0.01$ dex in the $\log_{10}(M_{\star}/M_{\odot}) = 8.5 - 10.0$ range, while the  M${\rm_{\star, SED-FAST}}$ are underestimated with respect to the M${\rm_{\star, intr}}$ by $0.17$ dex at $\log_{10}(M_{\star}/M_{\odot}) = 10.5$ at $z \simeq 1$. Similarly, in the right panel of Fig. \ref{AppendixFig} we show that the offset is $-0.15$ in the $\log_{10}(M_{\star}/M_{\odot}) = 8.5 - 10.0$ range. The derived stellar masses are underestimated by 0.25 dex at $\log_{10}(M_{\star}/M_{\odot}) = 10.5$ In conclusion, the stellar masses derived by FAST assuming an exponentially declining Star Formation Histrory (SFH) [$SFR = exp(-t/ \tau)$], the \citep{chabrier03} IMF, the \citet{Calzetti2000} dust attenuation law and a metallicity Z = 0.2 $Z_{\odot}$ are typically underestimated with respect the intrinsic values by 0.1 to 0.3 dex at $z \simeq 1-4$. \footnote{\citet{Camps2016} also demonstrated that masses inferred from the i band luminosity ${\rm L_{i}}$ and the g-i colour \citep{Cortese2012} from the EAGLE+SKIRT SEDs underestimate the stellar mass with respect the intrinsic values by $\simeq 0.25$ dex at $z \simeq 0$, pointing out differences between intrinsic and derived stellar masses.}

\begin{table*}
\centering
\resizebox{0.68\textwidth}{!}{%
  \begin{tabular}{ccccccccc}
    \hline \\
    & {\large Methodology} & {\Large $-0.5$} & {\Large $0$} & {\Large $0.5$} & {\Large $1.0$} & {\Large $1.5$}\\
     &  &  & & Offset  & (dex)        \\
    \hline \hline
    & SFR${\rm_{SED-FAST}}$, z = 4  & 0.02 & -0.15 & -0.26 & -0.48 & -0.85 \\
    & SFR${\rm_{SED-FAST}}$, z = 2  & -0.20 & -0.21 & -0.16 & -0.07 & -0.44  \\
    & SFR${\rm_{SED-FAST}}$, z = 1  & -0.03 & 0.06  & 0.19  & 0.24 & 0.11  \\
    \hline
    & SFR${\rm_{24 \mu m - Wuyts \, et \, al. \, 2008}}$, z = 4  & 0.49 & 0.37 & 0.40 & 0.40  & 0.16 \\
    & SFR${\rm_{24 \mu m - Wuyts \, et \, al. \, 2008}}$, z = 2  & 0.27 & 0.30 & 0.33 & 0.34 & 0.10 \\
    & SFR${\rm_{24 \mu m - Wuyts \, et \, al. \, 2008}}$, z = 1  & 0.15  &  0.17 & 0.17 & 0.13 & -0.14 \\
    \hline
     & SFR${\rm_{24, 70, 160 \mu m - r \, Dale \& Helou \, 2002}}$, z = 4   & 0.32 & 0.22 & 0.20 & 0.19 & 0.19 \\
    & SFR${\rm_{24, 70, 160 \mu m - r \, Dale \& Helou \, 2002}}$, z = 2   & 0.21  & 0.20  & 0.22 & 0.25 & 0.24 \\
    & SFR${\rm_{24, 70, 160 \mu m - r \, Dale \& Helou \, 2002}}$, z = 1   & 0.18  & 0.21 & 0.24 & 0.27 & 0.25 \\
    \hline
     & SFR${\rm_{250, 350, 500 \mu m - c \, Dale \& Helou \, 2014}}$, z = 4   & 0.32 & 0.20 & 0.20 & 0.21 & 0.27 \\
    & SFR${\rm_{250, 350, 500 \mu m - c \, Dale \& Helou \, 2014}}$, z = 2  & 0.20  & 0.20  & 0.23 & 0.30 & 0.40 \\ 
    & SFR${\rm_{250, 350, 500 \mu m - c \, Dale \& Helou \, 2014}}$, z = 1  & 0.15  & 0.18 & 0.24 & 0.36 & 0.50 \\
    \hline
    &  SFR${\rm_{UV+IRX-\beta}}$, z = 4  & 0.20 & 0.07 & 0.04 & -0.16 & -0.59 \\
    &  SFR${\rm_{UV+IRX-\beta}}$, z = 2  & 0.17 & 0.18 & 0.12  & -0.02  & -0.46 \\
    &  SFR${\rm_{UV+IRX-\beta}}$, z = 1  & 0.09 & 0.12 & 0.05 & -0.05  & -0.53 \\
\hline \hline
  \end{tabular}%
}
\caption{The offset in dex between the inferred and intrinsic SFRs at a fixed intrinsic SFR.}
\label{tab_stepsfrf5}
\end{table*}

In the bottom panels of Fig. \ref{AppendixFig} and table \ref{tab_stepsfrf5} (left panel $z \simeq 4$, middle panel $z \simeq 2$ and right panel $z \simeq 1$) we investigate the offset between the SFRs inferred from the indicators presented in section \ref{thecode2} and the intrinsic SFRs (${\rm SFR_{intr}}$). The blue dotted lines represents the offset between the SFR${\rm_{UV+IRX-\beta}}$ and intrinsic SFRs. At the lower SFR regime, the SFR${\rm_{UV+IRX-\beta}}$ are overestimated by $ \simeq 0.1-0.2$ dex. The authors suggested that the low SFR objects are passive galaxies with a low dust content, where the UV radiation emitted by the evolved star population is interpreted as the formation of new stars by the UV indicator. On the other hand, the derived SFRs are underestimated by up to -0.65 dex for high SFR objects. All the above are in agreement with the findings of \citet{Camps2016} for $z \simeq 0$. The UV-upturn (overestimation at low SFRs and underestimation at high SFRs) described above is evident as well in observations \citep{Brown2003}. The underestimation of the UV SFR with respect to other indicators in the high-SFR regime is also demonstrated in \citet{Katsianis2016} and \citet{Katsianis2017}.

The dark green dotted-dashed line represents the offset between SFR${\rm_{SED-FAST}}$ and SFR${\rm_{intr}}$. We demonstrate that the SFR${\rm_{SED-FAST}}$ are underpedicted at $z \simeq 4$ and $z \simeq 2$. The offset increases at high SFRs and can be up to $-0.6$ dex. This is in agreement with the findings of \citet{Conroy2013} who demonstrated that SED-based values, assuming a range of SFHs (including exponentially declining), metallicities, and dust attenuation laws, tend to be underpredicted, compared to a mixed UV+IR indicator. A range of other studies \citep{Brichram04,Salim2007} suggested as well that SFRs based on modeling UV-optical SEDs carry systematic uncertainties and underpredict the values with respect to UV+TIR indicators. We find that ${\rm SFR_{SED}}$ are underestimated with respect the intrinsic values at $z \simeq 2-4$ but at $z \simeq 1$ the derived ${\rm SFR_{SED}}$ are overestimated, especially for higher intrinsic SFRs.

The yellow dotted-dashed lines represents the offset between SFR${\rm_{24 \mu m - Wuyts \, et \, al. \, 2008}}$ and SFR${\rm_{intr}}$. For objects with intrinsic SFRs at the -0.5 to 1.0 regime SFRs are typically overestimated by 0.2-0.5 dex. This is in agreement with \citet{Rodighiero10}, \citet{Delooze2014} and \citet{Martis2019}. In contrast the derived $SFR_{24 \mu m}$ are underestimated for higher star forming objects. We note that the model assumed in the SKIRT post-process involves isotropically emitting star forming regions that may not represent the variations of the radiation field in these regions sufficiently. As a result, some fraction of the diffuse dust in the EAGLE galaxies may not be sufficiently heated, producing a lower 24${\rm \mu m}$ flux than expected \citep{Camps2016}. In addition, the 24${\rm \mu m}$ inferred SFRs could be underpredicted from the simulations if a significant fraction of photons from young stars is not successfully absorbed by dust \citep{Sklias2014,Hayward2014}.

The magenta solid line/red dashed line represents the offset between SFR${\rm_{24, 70, 160 \mu m - r \, Dale \& Helou \, 2002}}$/SFR${\rm_{250, 350, 500 \mu m - c \, Dale \& Helou \, 2014}}$ and SFR${\rm_{intr}}$. The methods overpredict SFRs by $ \simeq 0.1-0.5$ dex at $z \simeq 4$, while at $z \simeq 2$ and $ z \simeq 1$ the offset increases and is between $\simeq 0.2$ and $ \simeq 0.5 $ dex. This may be due to the fact that the emission from diffuse dust residing in the outskirts of the EAGLE+SKIRT galaxies is interpreted by the Total IR indicator as a sign of star formation \citep{Camps2016}, while the dust is heated by an evolved star population and not by newly born stars. The above IR contamination is also found in observations \citep{Helou2000,Bendo2015}.

\label{lastpage}
\end{document}